\RequirePackage{fix-cm}
\documentclass[smallextended]{svjour3}       
\smartqed  
\usepackage{graphicx}
 \journalname{Quantum Studies: Mathematics and Foundations}
\begin{document}

\title{Tunneling in energy eigenstates and  complex quantum trajectories  
}


\author{Kiran Mathew         \and
        Moncy V. John 
}


\institute{K. Mathew . M.V. John \at
              Department of Physics,  St. Thomas College, Kozhencherry 689641, Kerala, India
              Tel.: 91-9495128669\\
               \email{kiran007x@yahoo.co.in}           
           \and
          M.V. John \at
              Tel.: 91-9447059964\\
             \email{moncyjohn@yahoo.co.uk}
}

\date{Received: date / Accepted: date}

\maketitle

\begin{abstract}

Complex quantum trajectory approach, which arose from a modified de Broglie-Bohm interpretation of quantum mechanics, has attracted much attention in recent years. The exact complex  trajectories for the Eckart  potential barrier and the soft potential step, plotted in a previous work, show that more trajectories link the left and right regions of the barrier, when the energy is increased. In this paper, we   evaluate the reflection probability using a new ansatz based on these observations, as the ratio between the total probabilities of reflected  and  incident trajectories.  While doing this, we also put to test the complex-extended probability density previously postulated for these quantum trajectories. The new ansatz  is preferred since the evaluation is solely done  with the help of the complex-extended probability density along the imaginary direction and the trajectory pattern itself. The calculations are performed  for a rectangular potential barrier,  symmetric Eckart and Morse barriers, and a soft potential step. The  predictions are in perfect  agreement with the standard results for potentials such as the rectangular potential barrier. For the other  potentials,  there is very good agreement with standard results, but it is exact only for  low and high energies. For moderate energies, there are slight deviations. These deviations  result from the periodicity of the trajectory pattern  along the imaginary axis and have  a maximum value  only as much as $0.1 \%$ of the standard  value. Measurement of such  deviation shall provide an opportunity  to falsify  the ansatz.

\end{abstract}

\section{Introduction}

A modified de Broglie-Bohm (MdBB) approach to quantum mechanics was  proposed in 2001 \cite{mvj1} and the  complex quantum trajectories  envisaged in it  were drawn for several cases.  This  approach used the Schrodinger equation rewritten into a `quantum Hamilton-Jacobi equation'  \cite{qhj_eqn1,qhj_eqn2,qhj_eqn3,qhj_eqn4}  and employed  an equation of motion, both of which are similar to those used   in the  well-known  de Broglie-Bohm (dBB) formalism \cite{db1,valentini,bohm,holland}. But  the MdBB quantum trajectories contain all the information in the wave function and  are laid out  in a complex space (with  $x\equiv x_r+ix_i$ in one-dimension) and the scheme offers a different  interpretation of quantum mechanics \cite{sanz_comment,goldfarb_reply,sanz_review} than the dBB theory.  For the reason that the equation of motion used here is not based on Jacobi's theorem, just as the  dBB scheme, this too cannot be termed a `quantum Hamilton-Jacobi formalism'. (A candidate for a `quantum Hamilton-Jacobi formalism', which uses  Jacobi's theorem,  is  the Floyd-Faraggi-Matone (FFM) real trajectory representation \cite{floyd,faraggi,carroll}.) The MdBB complex trajectory formalism  could solve the  problem of zero  particle velocity for real wave function encountered  in the dBB approach and was found  capable of explaining many other quantum behaviour in a natural way \cite{yang1,yang2,sanz_local}. Earlier, the dBB  approach has helped to develop a computational method called quantum trajectory method (QTM) \cite{wyatt99}  to synthesize the Schrodinger wave function on the real axis.   In recent years, also the MdBB approach has been developed as a computational tool for this purpose \cite{sanz_review}. Bohmian mechanics with complex action  \cite{goldfarb} method is aimed at obtaining the wave function directly from the complex trajectories, as in QTM,  and was found more accurate. With the same spirit and by the same time, an alternative synthetic approach using complex trajectories was  developed by Chou and Wyatt \cite{wyatt1,wyatt2}, with applications in both bound states and scattering systems.   

A clear advantage of the MdBB complex representation  over other trajectory formalisms is that in it the Born probability distribution along the real line can be directly obtained from the velocity field  \cite{mvj_prob1}. It is seen that an exponential function involving the integral of the imaginary part of the MdBB velocity field  provides  this distribution in such a way  that on the real line, the more we are inside the closed trajectories, the larger is the probability density. No other trajectory representation  provides such an interpretation. Another feature  special  to  the complex trajectory approach is an extended probability density  in the complex plane. This issue was first addressed by Poirier \cite{poirier}, who proposed a complexified analytic probability density defined by  $\Psi^{\star}(x^{\star})\Psi(x)$, the product of the wave function and its generalized complex conjugate, and showed that it satisfies a complexified flux continuity equation in the complex plane.   An alternative  probability density of the form $\rho(x_r,x_i)$ in the complex space, which obeys a continuity equation  and which agrees with the Born probability along the real line  in most regions of the complex plane was proposed in \cite{mvj_prob1}.  This expression involves an integral of the imaginary part of the sum of kinetic and potential energies along the trajectories and is also obtained by direct solving of an appropriate continuity equation.  A problem with this scheme  is that in some regions of the complex plane, the distribution does  not agree with Born's probability along the real line. Chou and Wyatt \cite{wyatt_prob1} have pointed out that the density proposed by Poirier  mentioned above does not match the node structure of the wave function. In place of it, they have proposed an extended Born probability density $\Psi^{\star}(x)\Psi(x)$ in the complex space \cite{wyatt_prob1}. The authors could derive the Eulerian rate equation for this density  \cite{wyatt_prob2},  though as pointed out by them, this distribution does not obey a complex version continuity equation due to its nonanalyticity.
Later, it was observed that in the regions left out in Ref. \cite{mvj_prob1}, a trajectory integral of the above type (which involves only the kinetic energy in place of the sum of kinetic and potential energies), can agree with the Born probability density  \cite{mvj_prob2}.   This latter density was identified to be the same as the complex-extended  $\Psi^{\star}(x)\Psi(x)$   proposed earlier \cite{wyatt_prob1}. However, we note that in regions where the potential $V=0$, the two  extended probability densities are described by identical expressions.

  In this paper, we  put to test the   complex-extended probability distribution described  above \cite{wyatt_prob1,wyatt_prob2,mvj_prob2}. This is   by using it and the trajectory pattern to explain the quantum phenomenon of  tunneling  in energy eigenstates. Tunneling of particles in eigenstates is an interesting problem in itself and is well-studied in standard quantum mechanics \cite{flugge,zafar}. But  in  the literature on dBB quantum theory,  we are unable to find any demonstration of tunneling in energy eigenstates.  It appears that  in this scheme,  for such eigenstates, all trajectories which approach the barrier get transmitted \cite{norsen}. The  FFM trajectory representation describes in the incident region the creation and annihilation of pairs of trajectories going in opposite directions \cite{floyd_tunnel}, thus bringing in pair production in a nonrelativistic theory. On the other hand,  in the complex MdBB method, the trajectory pattern in the complex plane,  as plotted  for a rectangular potential step  \cite{mvj1}, shows clearly how certain trajectories in the incident region correspond to particles moving towards the barrier and get transmitted, whereas some others  correspond to  particles reflected back from it. For  realistic, smooth potentials such as the symmetric Eckart potential and the soft potential step, the complex trajectory patterns   drawn for eigenstates by Chou and Wyatt  \cite{wyatt_eckart,wyatt_eckart1}  explicitly show  this behaviour. But we note that   the transmission and reflection probabilities are so far evaluated only for  wave packets  \cite{goldfarb,wyatt1,rowland}. The procedure adopted in those cases is to start with a wave packet impinging on the potential barrier and then to evaluate the Born probability on the other side of the barrier, along the real line, after a sufficiently long time. In their work,  the complex quantum trajectory method is used  as just a computational tool as in the dBB case, since only the probability distribution on the real line   plays any role in it. In the present attempt, we assume the extended probability density \cite{wyatt_prob1,mvj_prob2}  for energy eigenstates and then, using the complex trajectory pattern,  compute the probabilities corresponding to the trajectories which actually tunnel or get reflected  from the barrier. 
  
  In Sec. II, we shall state our  ansatz and evaluate the reflection probability for a  rectangular potential barrier. In Sec. III, 
a continuous potential barrier, with the Eckart and Morse barriers as special cases is considered for this purpose. In Sec. IV, reflection probability for a soft potential step is evaluated. The probabilities we obtain using the new ansatz are exact for the rectangular potentials. For continuous potentials, they are in   excellent agreement with the corresponding results in standard quantum mechanics for low and high energies, but  slight deviations are noted for  moderate energies.  The last section comprises a discussion of these results and  our conclusions.
 
\section{Tunneling  a Rectangular Potential Barrier}

In the simplest one-dimensional version of tunneling, we have the superposition of an incoming  and a reflected wave on one side  of a square potential barrier ($V=V_0$, for $|x| < a$ and $V=0$, for $|x| > a$), given by

\begin{equation}
\psi (x) = A e^{ikx}+B e^{-ikx}, \label{eq:soln_left} 
\end{equation}
where $k=\sqrt{2mE}/\hbar$. Inside the barrier,  the solution is  of the form

\begin{equation}
\psi(x)= C e^{-\kappa x} + De^{+\kappa x}.\label{eq:soln_mid} 
\end{equation}
Here  $\kappa =\sqrt{2m(V_0-E)}/\hbar$. Assuming the particle to be incident from left ($x<-a$),  we have only 

\begin{equation}
\psi (x) = F e^{ikx}, \label{eq:soln_right} 
\end{equation}
as the wave function on the right side ($x>a$). In the standard case, we evaluate the tunneling probability  as    $T(k) \equiv |F| ^2/ |A|^2$ and the reflection probability as     $R(k) \equiv  |B|^2/|A|^2$.  The matching conditions, namely, the continuity of the wave function and its space derivative on the boundaries at $x=\pm a$, help to evaluate the tunneling probability as

\begin{equation}
T(k)= \frac{1}{1+[1+(\epsilon^2/4)]\sinh^2(2\kappa a)}, \label{eq:rect_tunnel_std}
\end{equation}
where $\epsilon = (\kappa /k)-(k/\kappa)$.    The above wave functions $\psi(x)$  constitute an  energy eigenstate of the particle, with a fixed value for  its energy. Since this is a time-independent problem with definite energy, we have $\Delta E =0$ and  it is meaningless to use the uncertainty relation $\Delta E \Delta t \geq \hbar$ to explain tunneling. As mentioned above, a more general description of tunneling phenomenon   in standard quantum mechanics involves  a wave packet  impinging on such a barrier. For evaluating the tunneling probability in this case,  one evaluates the total probability on the other side of the barrier,  a long time after the packet reaches it. But we may note that also in this approach,  there is a long-standing debate over whether it leads to  superluminal speeds for the particles as the width of the barrier exceeds  certain value, a predicted  phenomenon which is  referred to as  Hartmann effect \cite{hartman}.  In
the  case of a square potential barrier, even for an energy eigenstate, the tunneling time is given by an expression
that approaches a constant for thick barriers, implying an
issue concerning superluminal propagation, even though physical causality is not violated \cite{davies}. Hence  it is desirable to look forward to alternative formalisms, such as quantum trajectory representations, for more insight on quantum tunneling in energy eigenstates.

First we consider the dBB formalism.  This scheme adopts the  standard Born  probability axiom so that the trajectories do not have any particular role in the evaluation of reflection and transmission probabilities. It is easy to see that in the incident region where $\psi $ is given by Eq. (\ref{eq:soln_left}), the velocity of particles  given by  de Broglie's equation of motion, 

\begin{equation}
\dot{x}=\frac{\hbar}{2im}\frac{\left( \psi^{\star}\frac{\partial \psi}{\partial x} - \frac{\partial \psi^{\star}}{\partial x}\psi\right)}{\psi^{\star}\psi}, \label{eq:dBeqn_mtn}
\end{equation}
is always positive, assuming $|A| > |B|$. Thus according to this theory, in this region, we have particles traveling only towards the barrier. This means that  the dBB scheme disagrees with  standard quantum mechanics in the  tunneling of particles in energy eigenstates. Thus, in this scheme, by accepting  Born's probability rule as an axiom at the outset, the discrepancy is  swept under the rug.  Continuing with the use the standard probability axiom without relating probability to the trajectories gives the wrong impression that dBB agrees with standard quantum mechanics in the prediction of tunneling in eigenstates.

\begin{figure*}

\resizebox {1.0 \textwidth} {0.4 \textheight }{\includegraphics {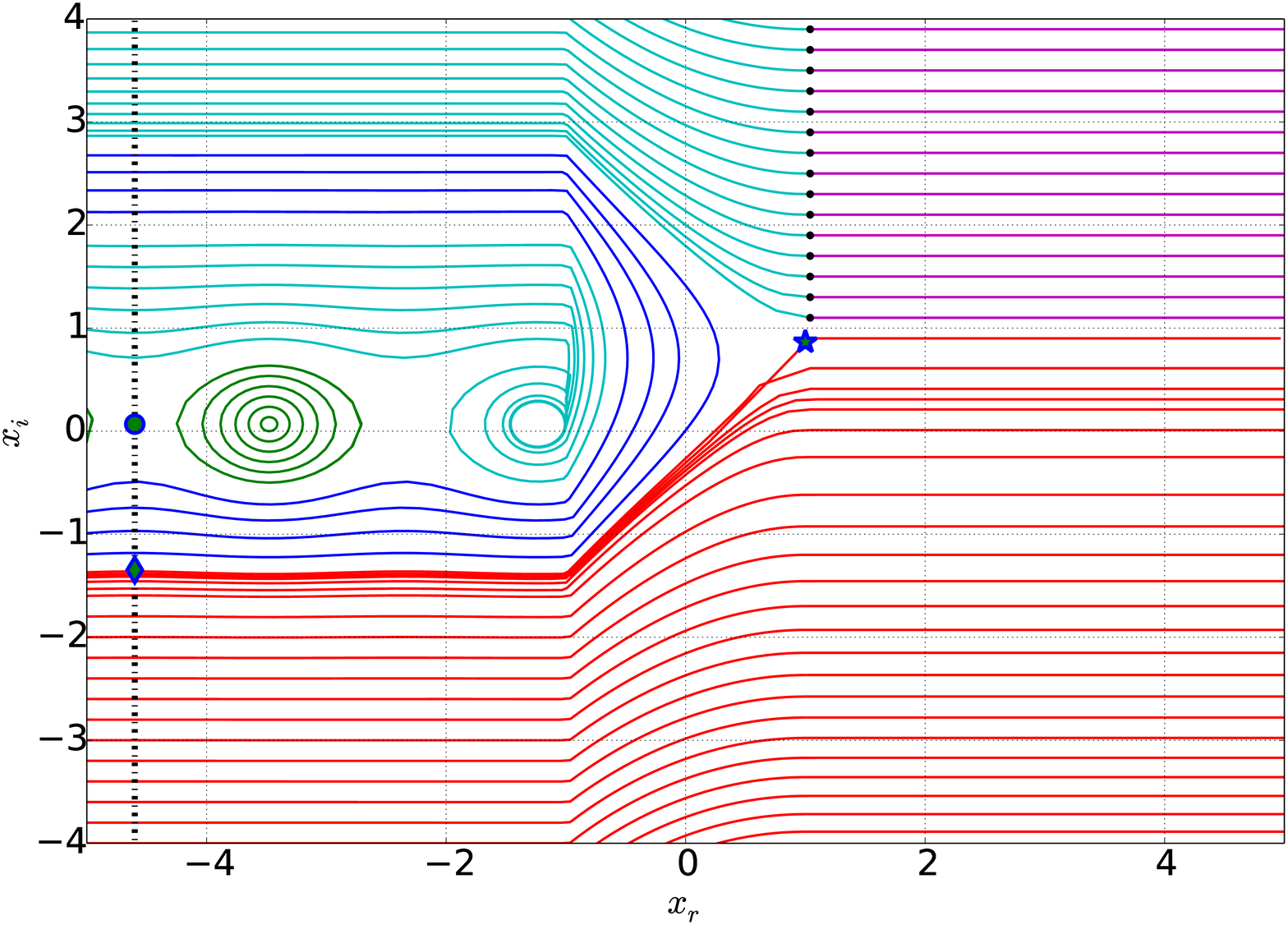}}
\caption{MdBB complex quantum trajectories for a rectangular potential barrier. The three regions, whose common  boundaries are at $x_r=-1$ and $x_r=+1$, have their characteristic trajectories obtained by integrating  the equation of motion (\ref{eq:eqn_motn}) using the respective wave functions (\ref{eq:soln_left})-(\ref{eq:soln_right}). These trajectories are drawn for an energy E=0.98. Color codes are as follows: Trajectories incident from bottom left and which tunnel through the barrier (red), similar trajectories which are reflected back (blue), closed loops (green), trajectories emanating from repellers and moving towards either left or right (cyan or magenta, in the  respective order). The points $\alpha +i \beta =-4.5998+0.0681 \; i $, $\gamma=-4.5998-1.347 \; i $ and $\delta =1+0.865\; i $ are marked with blue symbols $\bullet$, $\diamond$, and $\star$, respectively.}  \label{fig:rect_tunnel}
 
\end{figure*}

On the other hand, the MdBB approach helps to predict the tunneling probability on the basis of the  properties  of the trajectories itself and using the expression for the complex-extended probability density. The velocity field  given by the MdBB equation of motion is

\begin{equation}
\dot{x}=\frac{\hbar}{im}\frac{1}{\psi}\frac{\partial \psi}{\partial x}, \label{eq:eqn_motn}
\end{equation}
which is  defined in the complex plane.  Integrating this, we get the complex quantum trajectories. For a rectangular potential barrier, the trajectories are as shown in Fig. \ref{fig:rect_tunnel}.  The matching conditions on $\psi$ at the points $x=-a$ and $x=+a$ ($a$ is real), with which the coefficients $A$, $B$, etc. in equations (\ref{eq:soln_left})-(\ref{eq:soln_right}) are evaluated, can be understood as the conditions for   the continuity of the MdBB velocity field (\ref{eq:eqn_motn}) at these points. At other points along these boundaries at $x_r=\pm a$, for the rectangular potential barrier, the velocity field may not be continuous, though the trajectories can remain continuous. Another special feature to be noted is that there are trajectories emanating from some points along these boundaries (such as those  points at $x_r=+a$, for $x_i >\delta$ shown in figure). These trajectories  move in opposite directions and proceed either to the left or right sides. (Existence of such  points, which may be called repellers \cite{wyatt_eckart,wyatt_eckart1}, is a general feature of complex trajectories in scattering problems. In the following, we too demonstrate this in specific examples of realistic potentials.)

 Next, let us consider a `pole' for $\dot{x}$ (where $\psi =0$) in the complex plane,  on the incident side \cite{wyatt1}.  Let this point be denoted as $\alpha + i \beta$. One can see that along a line $x_r=\alpha$ parallel to the imaginary axis, $\dot{x}_r >0$  when $x_i<\beta$  and  $\dot{x}_r<0$  for  $x_i>\beta$. Thus when the equation of motion is integrated, in the $x_r<-a$ region, we obtain  trajectories of incident particles crossing this line  towards the barrier, for $x_i<\beta$. Similarly,  trajectories for reflected particles cross this line in the reverse direction, for $x_i > \beta$.  On the other ($x_r>a$) side of the barrier, we have only the transmitted trajectories for all values of $x_i$. It is seen that some of these trajectories emanate from points (which are repellers) on the boundary line $x_r=a$, with $x_i>\delta$, as shown in Fig. \ref{fig:rect_tunnel}. The various types of trajectories obtained, for an energy $E=0.98$, are drawn with different colors. In this paper, we show that the existence of such intuitively clear trajectory picture  helps to  evaluate the reflection and  transmission probabilities.

\begin{figure} 
 \resizebox {0.4 \textwidth} {0.3 \textheight }
{\includegraphics {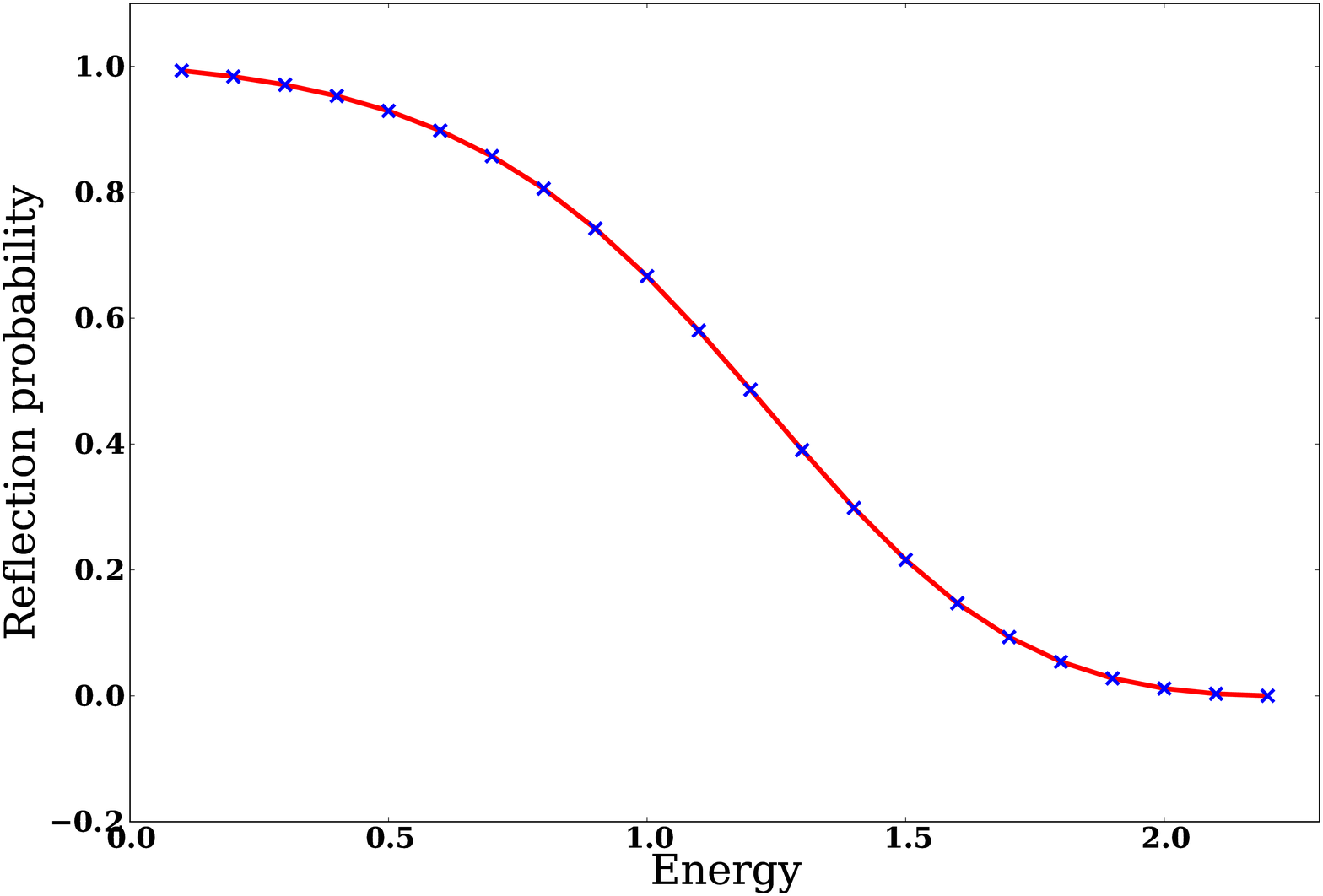}} 
\caption{Reflection probability  for a rectangular potential barrier, for the energy values $0.1\leq E\leq 2.2$. Blue x marks indicate values obtained using ansatz (\ref{eq:ansatz_2Lambda}) and red continuous line indicates values obtained in standard quantum mechanics.}  \label{fig:rect_prob}
  \end{figure}

 In the original paper of MdBB  formulation \cite{mvj1},   it was proposed that the  probability axiom apply  in  the same manner as that in the dBB approach; i.e., by obeying  Born's probability density on the real line. As stated above,   this  density is generalized to the complex-extended probability density  \cite{mvj_prob1,wyatt_prob1,mvj_prob2}.  The present work attempts to show that one can evaluate the tunneling and reflection probabilities by  making direct use of  the extended probability density, together with a new definition of reflection probability based on the  properties  of the trajectories. This new definition   involves the ratio of two integrals, both of which are evaluated along the above-mentioned  line $x_r=\alpha$ ($\alpha$ real and negative), passing through a pole for $\dot{x}$ (where $\psi=0$)  and parallel to the  imaginary axis. The integrand in both cases is the extended probability density $\psi^{\star}(x)\psi (x)$.  The  integral in the numerator gives  the total probability of reflected trajectories, obtained by integrating $\psi^{\star}(x)\psi (x)$   over them (i.e., for the part of the line with  $x_i>\beta$)  and the integral in the denominator is the total probability of incident trajectories,  obtained by integrating this density over all such trajectories (i.e., for  $x_i<\beta$).  In the present case of the rectangular potential barrier, the $\psi^{\star}\psi$ probability density diverges as $x_i \rightarrow \pm \infty$. Hence, we must take the integrals from  $x_i=\beta$ to $+\Lambda$ and from $x_i=- \Lambda$ to $\beta$, respectively, and then take the limit $\Lambda \rightarrow \infty$. It is seen that the value converges in the limit. Thus the new ansatz for the reflection probability is 

\begin{equation}
 R(k) =\lim_{\Lambda \to\infty}\frac{\int_{\beta}^{\Lambda} \psi^{\star}\psi \; dx_i}{\int_{-\Lambda}^{\beta} \psi^{\star}\psi \; dx_i}. \label{eq:ansatz_2Lambda}
\end{equation}
 The reflection probability evaluated numerically using this ansatz is shown in Fig. \ref{fig:rect_prob}, which is in good agreement with standard result $R(k) = 1-T(k)$, where $T(k)$ is given by Eq. (\ref{eq:rect_tunnel_std}). When  $\Lambda $ is finite, there are some deviations for moderate energies ($E\approx V_0$), but we notice that such deviations disappear for $\Lambda \rightarrow \infty$. 

Though we have evaluated the above expression numerically, it is easy to see analytically, by substituting (\ref{eq:soln_left}) in (\ref{eq:ansatz_2Lambda}), that $R(k) \rightarrow  |B|^2/|A|^2 $ in the limit  $\Lambda \rightarrow \infty$. The line of integration is parallel to the imaginary axis and passes through a pole of the velocity field so that it avoids the separatrix  surrounding the web of nodes. For   energy $E\rightarrow 0$, the value of $\beta$ is close to zero and the trajectory pattern is symmetric about the real line. So the above ratio tends to the value unity, equal to that in the standard calculation. For  energies $E>0$ too,  the ratio  tends to the value  $|B|^2/|A|^2 $   since the integrals are not sensitive to the relatively small values of $\beta$ \cite{wyatt1} when compared to large values of $\Lambda$ as $\Lambda \rightarrow \infty$.

The primary motivation for choosing the expression (\ref{eq:ansatz_2Lambda}) as the reflection probability is that it is the ratio between the probability for the reflected trajectories and that for the incident trajectories, evaluated along the line $x_r=\alpha $ on the incident side of the complex plane. The fact that it is able to reproduce the exact values obtained in standard calculations encourages us to use a similar expression for reflection probability, also in the case of  realistic potentials. This is  discussed in the next sections.

\section{Tunneling the Eckart and Morse Barriers}

We may   note that  the rectangular potential barrier is just an idealization.  Therefore it is desirable to look  into  the case of more realistic potentials, such as the symmetric Eckart and Morse potential barriers or the soft potential step. Before that, let us consider an arbitrary potential barrier whose   peak along the real axis is  at $x=0$. At large distances from the barrier ($x_r \rightarrow \pm \infty$),  let the potential  be such that  $V\rightarrow 0$. The time-independent Schrodinger equation   has  asymptotic solutions of the form found in  Eqs. (\ref{eq:soln_left}) and (\ref{eq:soln_right}). (But it may be noted that  in the MdBB approach, we have  their analytic continuations, with $x=x_r+ix_i$, as solutions.) In the following, we consider a class of such  potentials, which exhibit  periodicity along the imaginary line.  With the corresponding period  denoted as $2L$, we  modify our  ansatz for reflection probability  as

\begin{equation}
 R(k) =\frac{\int_{\beta}^L \psi^{\star}\psi \; dx_i}{\int_{-L}^{\beta} \psi^{\star}\psi \; dx_i}. \label{eq:ansatz}
\end{equation}
The value of $x_r=\alpha$ (on the line along which the integrals are taken)  must be sufficiently large and negative. Note that here also, the numerator  and the denominator represent the total probability for the reflected and incident trajectories, respectively.

  The Eckart and Morse barriers are special cases of a potential barrier in one-dimension studied by Ahmed \cite{zafar}, of the form

\begin{equation}
V(x) = V_0 \left\lbrace 1- \left[ \frac{1-\exp(x/a)}{ 1+c\exp(x/a)} \right]^2 \right\rbrace . \label{eq:ahmeds_potential}
\end{equation}
 When $c=0$, it is the Morse barrier and when $c=1$, this gives the symmetric Eckart barrier. Using the transformation $z=-c\exp(x/a)$, and using $b^2=1/c^2$, $\Delta = \hbar^2/(2ma^2)$, $f^2=E/\Delta$, $q^2=V_0/\Delta$, and 
 
 $$s=\sqrt{f^2+(b^2-1)q^2},$$
 
  $$g=\sqrt{q^2(b+1)^2-(1/4)},$$

$$k=\sqrt{2mE}/\hbar,$$
 and 
  
   $$k^{\prime} =\sqrt{2m[E+(b^2-1)V_0]}/\hbar,$$
  the exact solutions for the   Schrodinger equation are obtained  \cite{zafar}, in terms of the hypergeometric functions. One such solution is

\begin{eqnarray}
\psi_i = (-1)^{-if} (2f)^{-1/2} \exp(-\pi f)z^{if}(1-z)^{(1/2)-ig}{}_2F_1[(\frac{1}{2}+if-ig-is),\nonumber \\ 
 (\frac{1}{2}+if-ig+is), (1+2if);z]\exp(-iEt/\hbar), \label{eq:psieckart_inc}
\end{eqnarray}
which behaves as $\exp[i(kx-Et/\hbar))]$ as $x\rightarrow -\infty$, denoting a wave function $\psi_i$ incident on the barrier. Changing $k$ to $-k$  in this gives another solution

\begin{eqnarray}
\psi_{\rho} = (-1)^{if} (2f)^{-1/2} \exp(\pi f)z^{-if}(1-z)^{(1/2)-ig}{}_2F_1[(\frac{1}{2}-if-ig-is),\nonumber \\ 
 (\frac{1}{2}-if-ig+is), (1-2if);z]\exp(-iEt/\hbar).\label{eq:psieckart_refl}
\end{eqnarray}
This represents an oppositely traveling, reflected wave.  Another useful solution is

\begin{eqnarray}
\psi_{\tau} = (-1)^{if} (2s)^{-1/2} \exp(-\pi f)z^{-if}(1-z)^{is+if}{}_2F_1[(\frac{1}{2}-if-ig-is),\nonumber \\ 
 (\frac{1}{2}-if+ig-is), (1-2is);(1-z)^{-1}]\exp(-iEt/\hbar), \label{eq:psieckart_trans}
\end{eqnarray}
which in the limit $x\rightarrow +\infty$ denotes a wave $\exp[i(k^{\prime}x-Et/\hbar)]$ being transmitted through the barrier. Assuming a desirable relation among  $\psi_i$, $\psi_{\rho}$, and $\psi_{\tau}$  \cite{zafar}, namely,

\begin{equation}
\psi_i +\rho (k) \psi_{\rho} = \tau (k)\psi_{\tau}, \label{eq:zafar_assmp}
\end{equation}
one can evaluate the reflection  and transmission  amplitudes $\rho(k)$ and $\tau(k)$, respectively. The corresponding probabilities are found as $R(k)= \rho^*(k)\rho(k)$ and  $T(k) = \tau^*(k)\tau(k)$. Explicitly, we get

\begin{equation}
R(k) =  \frac{\cosh[\pi (f+g-s)]\cosh[\pi(f-g- s)]}{\cosh([\pi (f+s+g)]\cosh[\pi(f+s-g)]}, \label{eq:zafar_R}
\end{equation}

and 
 
  \begin{equation}
T(k) = \frac{\sinh(2\pi f)\sinh(2\pi s)}{\cosh([\pi (f+s+g)]\cosh[\pi(f+s-g)]}.\label{eq:zafar_T}
\end{equation}

We note that the derivation of these expressions for $R(k)$ and $T(k)$  depends on the assumption (\ref{eq:zafar_assmp}).   But as can be seen from \cite{wyatt_eckart,wyatt_eckart1},  the MdBB trajectory pattern is obtained with $\psi_{\tau}$ alone and   this formalism clearly distinguishes incoming, reflected and transmitted trajectories. In the present paper, we show that also the above probabilities are obtainable from   $\psi_{\tau}$ alone, with the help of ansatz  (\ref{eq:ansatz}).

The analytic continuations into the complex plane of the Eckart, Morse and other potentials discussed below in this paper have periodicity along the imaginary axis. Since they contain a factor $\exp(x/a)$, these potentials given by  Ahmed's expression (\ref{eq:ahmeds_potential}) have the same value for $x=x_r+i x_i$ and $x=x_r+i (x_i+2\pi a)$. Furthermore, the complex powers are calculated using the principal values of the argument of the complex number $x$ by specifying $-\pi < \arg(x)\leq +\pi$, and this gives rise to a discontinuity of the velocity field along $x_i=\pm \pi a i$, $\pm 3 \pi a i$, ... for $x_r<0$. The trajectory pattern so obtained has the same periodicity as that of the potentials. This is explicitly shown further  below, while plotting trajectories in specific cases.

  \begin{figure*} 
\resizebox {0.3 \textwidth} {0.3 \textheight }{\includegraphics {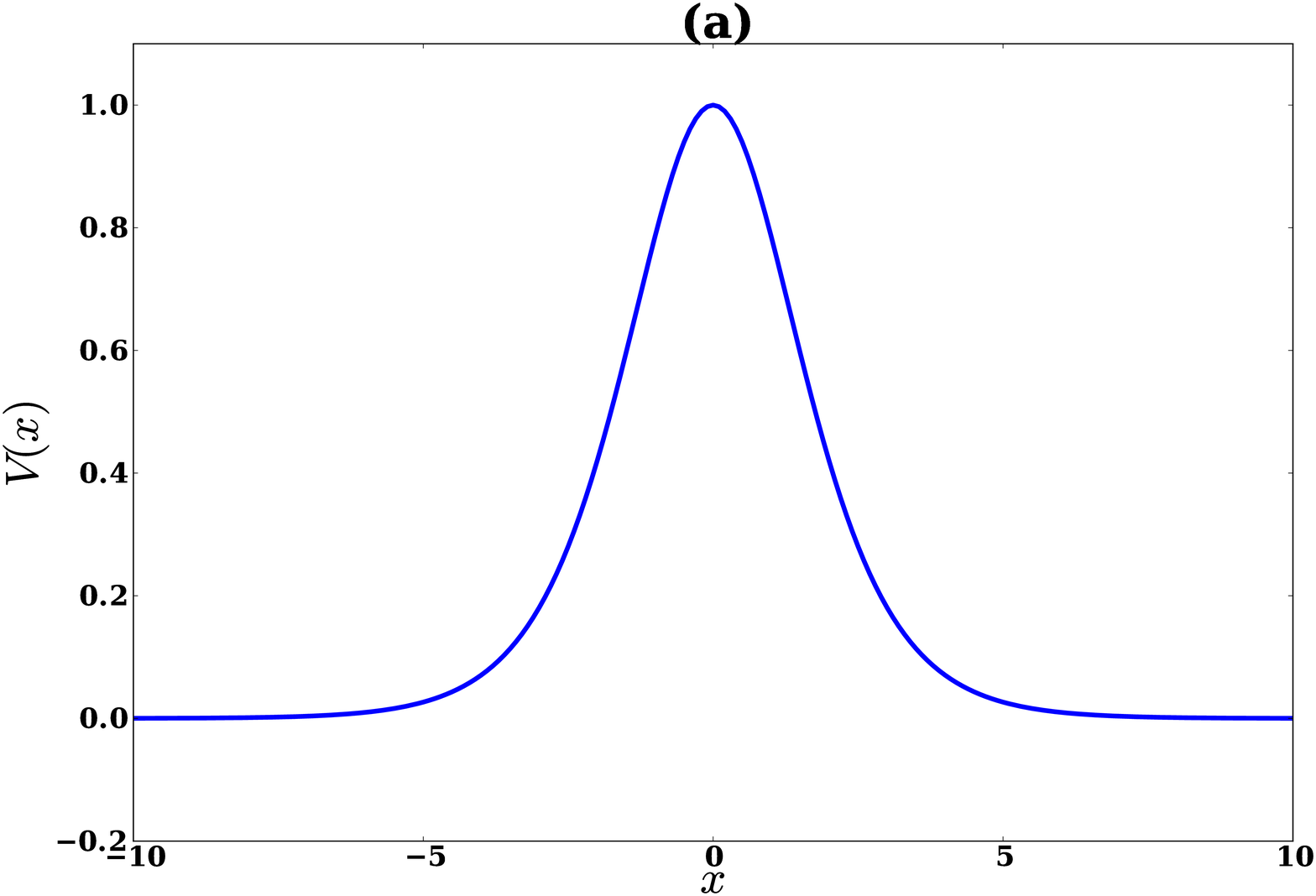}},\resizebox {0.3 \textwidth} {0.3 \textheight }{\includegraphics {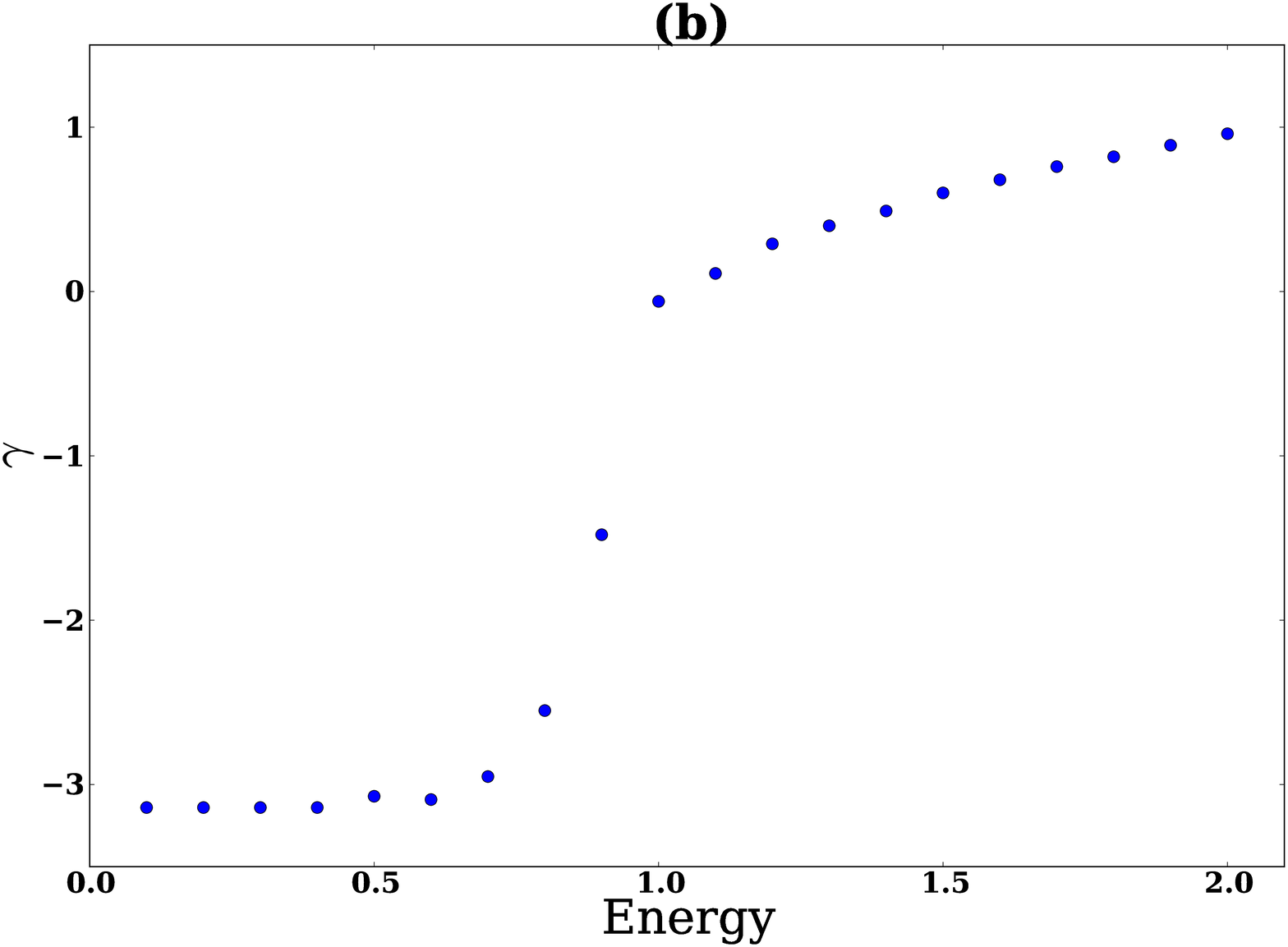}},\resizebox {0.3 \textwidth} {0.3 \textheight }{\includegraphics {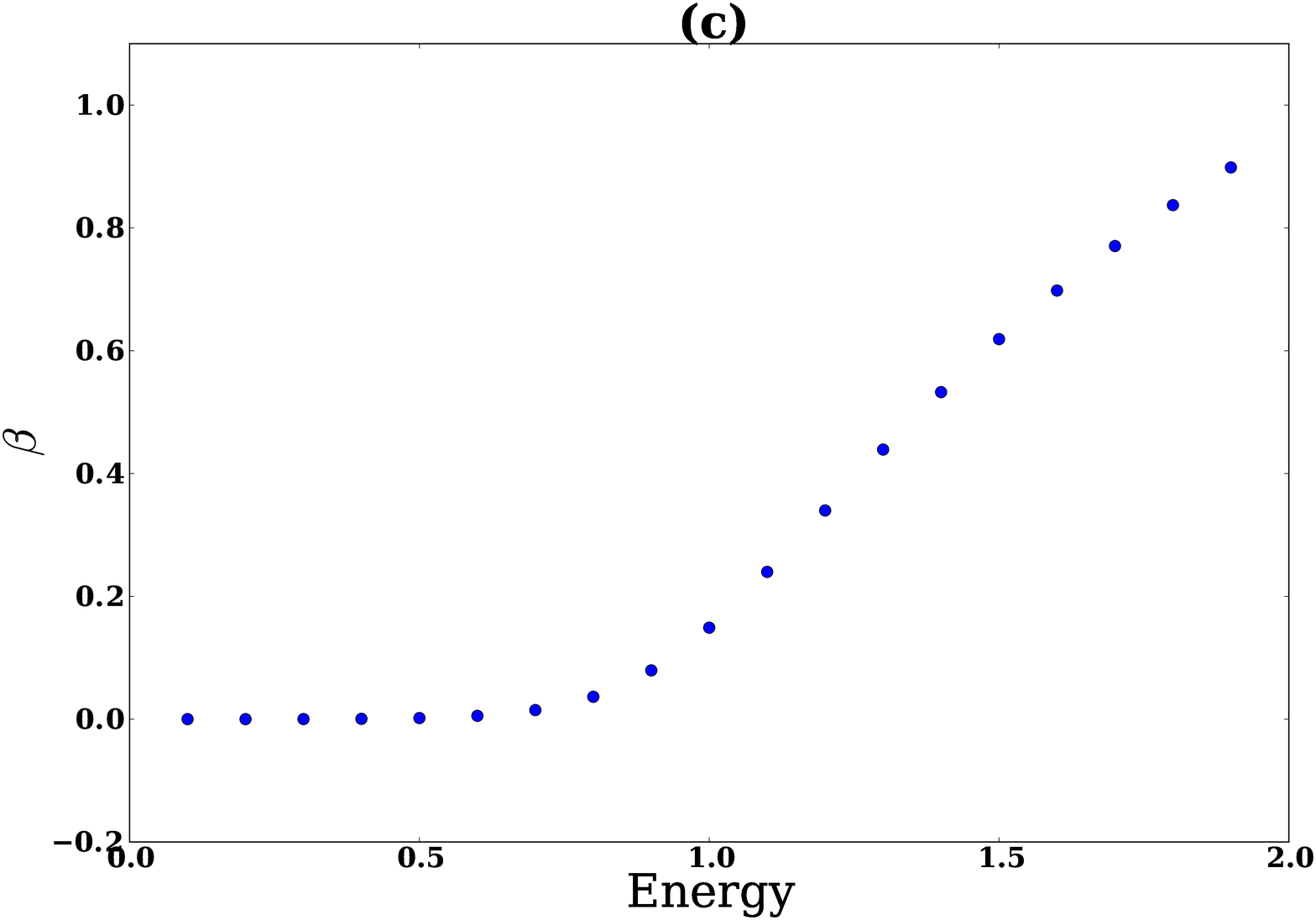}}
\caption{(a) Symmetric Eckart potential (b) Variation of the parameter $\gamma$ with energy for symmetric Eckart potential (c) Variation of the parameter $\beta$ with energy for symmetric Eckart potential  }  \label{fig:Eck_params}
  \end{figure*}

\begin{figure*}  
\resizebox {0.5 \textwidth} {0.3 \textheight }{\includegraphics {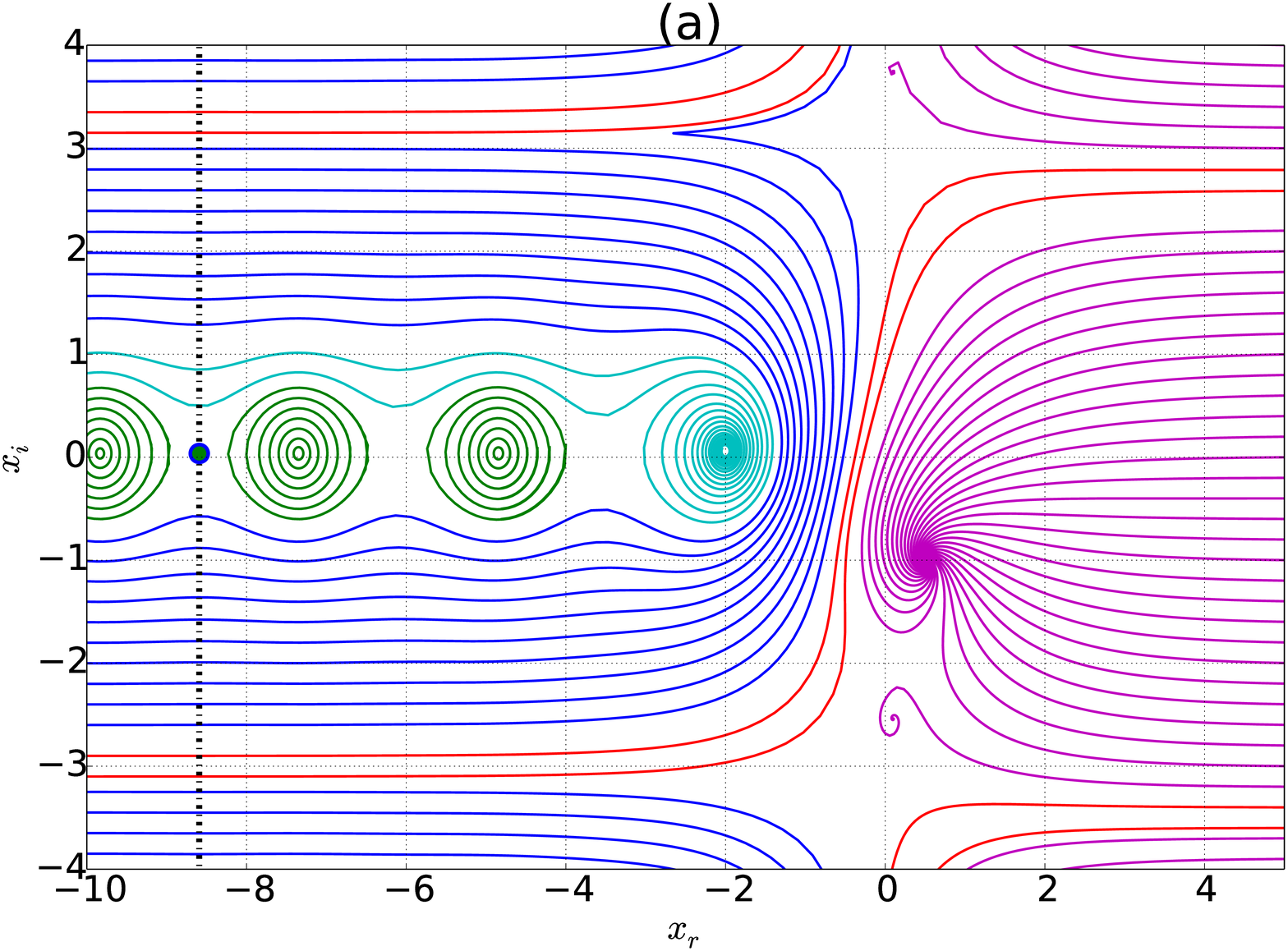}},\resizebox {0.5 \textwidth} {0.3 \textheight }{\includegraphics {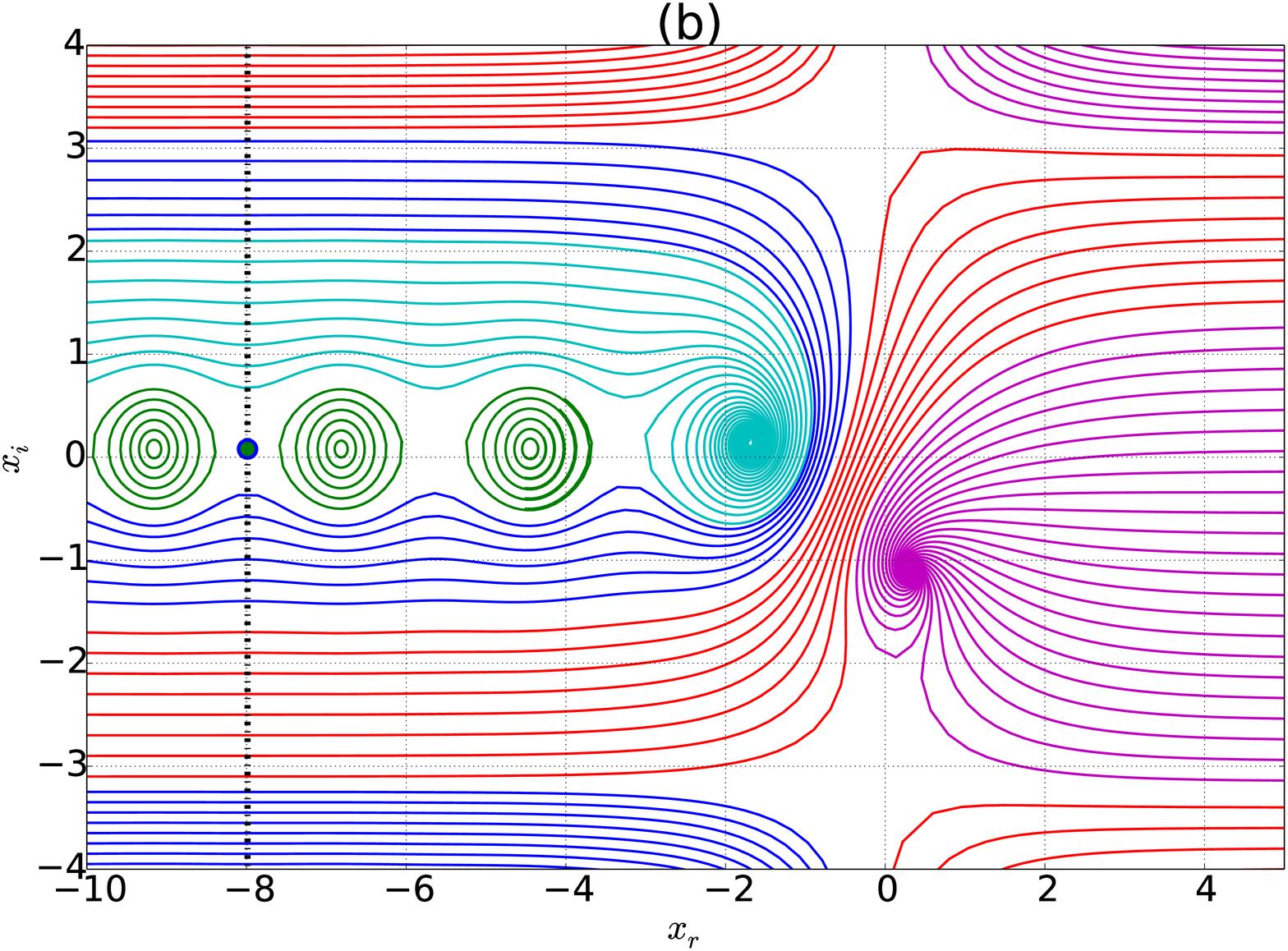}}, \\ \resizebox {0.5 \textwidth} {0.3 \textheight }{\includegraphics {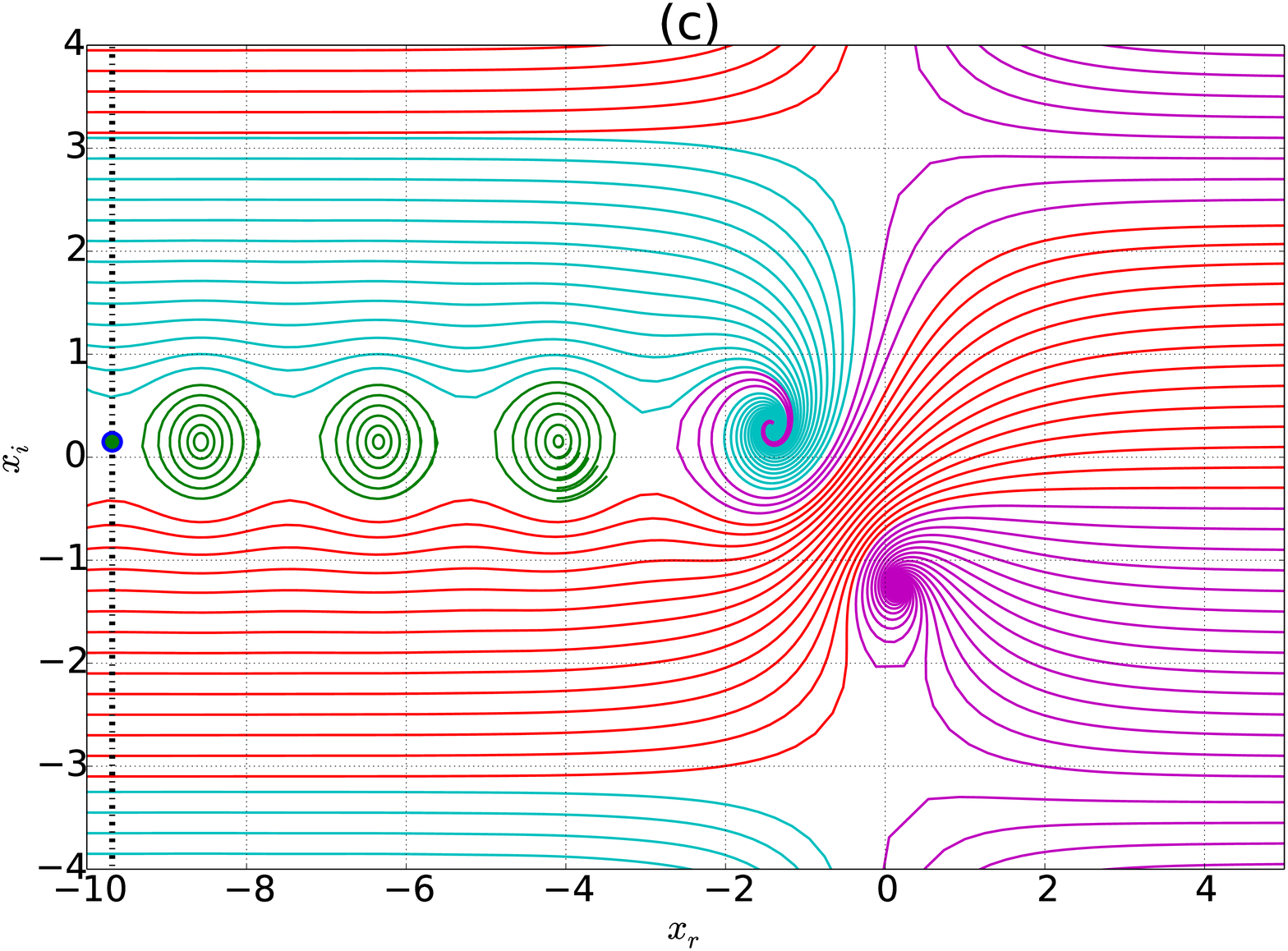}},\resizebox {0.5 \textwidth} {0.3 \textheight }{\includegraphics {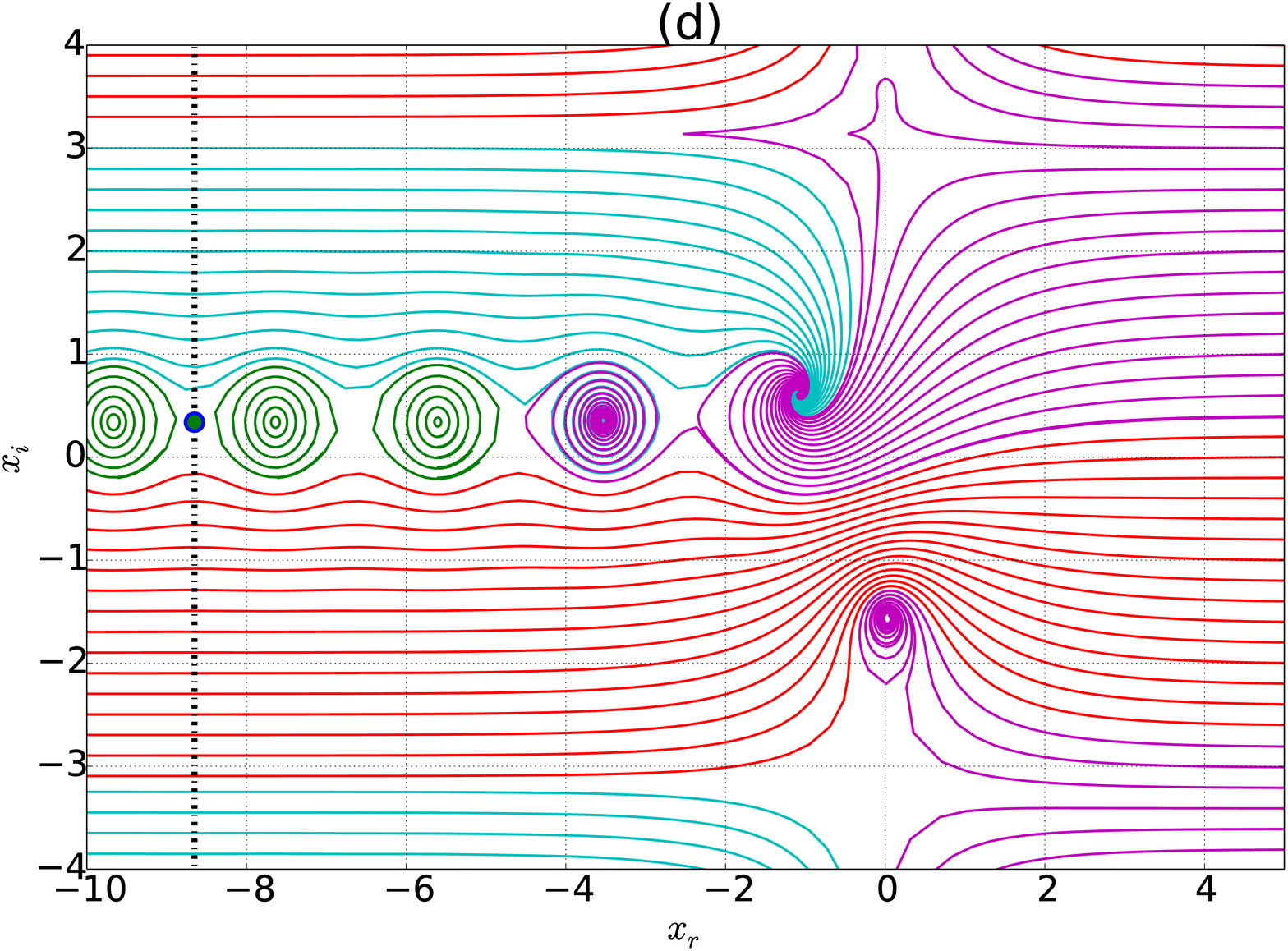}} 
\caption{Complex quantum trajectories for symmetric Eckart potential, for the energy values (a) $E=0.8$, (b) $E=0.9$ (c) $E=1.0$ and (d) $E=1.2$. Color codes are as in Fig. 1. In each panel, the points $\alpha +i\beta$, whose numerical values  are $ -8.5918		+0.036484\; i$, $-7.98707+	0.079476\; i$, $-9.68208	+0.149013\; i$, $-8.65166+	0.339926\; i$, respectively,  are marked with $\bullet$ symbols. }  \label{fig:Eck_trajs}
  \end{figure*}

We choose the symmetric Eckart potential  case [Fig. \ref{fig:Eck_params}(a)] by putting $c=1$ in the above equations and plot the   complex trajectories   using $\psi_{\tau}$ in the equation of motion (\ref{eq:eqn_motn}). This is repeated  for various energies $E$. We fix $V_0=1$,   $a=1 $ and $\Delta = \hbar^2/(2ma^2)=1/2$. The trajectory pattern for various energies drawn in Fig. \ref{fig:Eck_trajs} are the same as those obtained earlier by Chou and Wyatt \cite{wyatt_eckart,wyatt_eckart1}. There is a periodicity in the pattern along the imaginary direction, with period $2\pi$ and hence we need only to consider the interval [$-\pi$,$\pi$] in this case. It may be noted that when $E\ll V_0$, very few trajectories tunnel and all the rest get reflected. But as the energy increases, more and more trajectories tunnel. As in the previous section, let us consider a line $x_r=\alpha$, which is parallel to the imaginary axis and which passes through a pole for the velocity field (where  $\psi =0$) at $\alpha +i \beta$.  Here  $\alpha$ must be large and negative. We denote the value of $x_i$, corresponding to the top-most tunneling trajectory (among the incident trajectories) when it  crosses this line, as $\gamma$.  As in the rectangular potential barrier case, the top-most incident trajectory while crossing this line hits the pole for the velocity field, where  $x_i=\beta$. Here, it must be noted that the values of $\alpha$, $ \beta$ and $\gamma$ vary with energy. Graphs showing the variation of $\gamma$ and $\beta$ with energy are given in Figs. \ref{fig:Eck_params}(b) and \ref{fig:Eck_params}(c), respectively.

\begin{figure} 
\resizebox {0.4 \textwidth} {0.3 \textheight }  
{\includegraphics {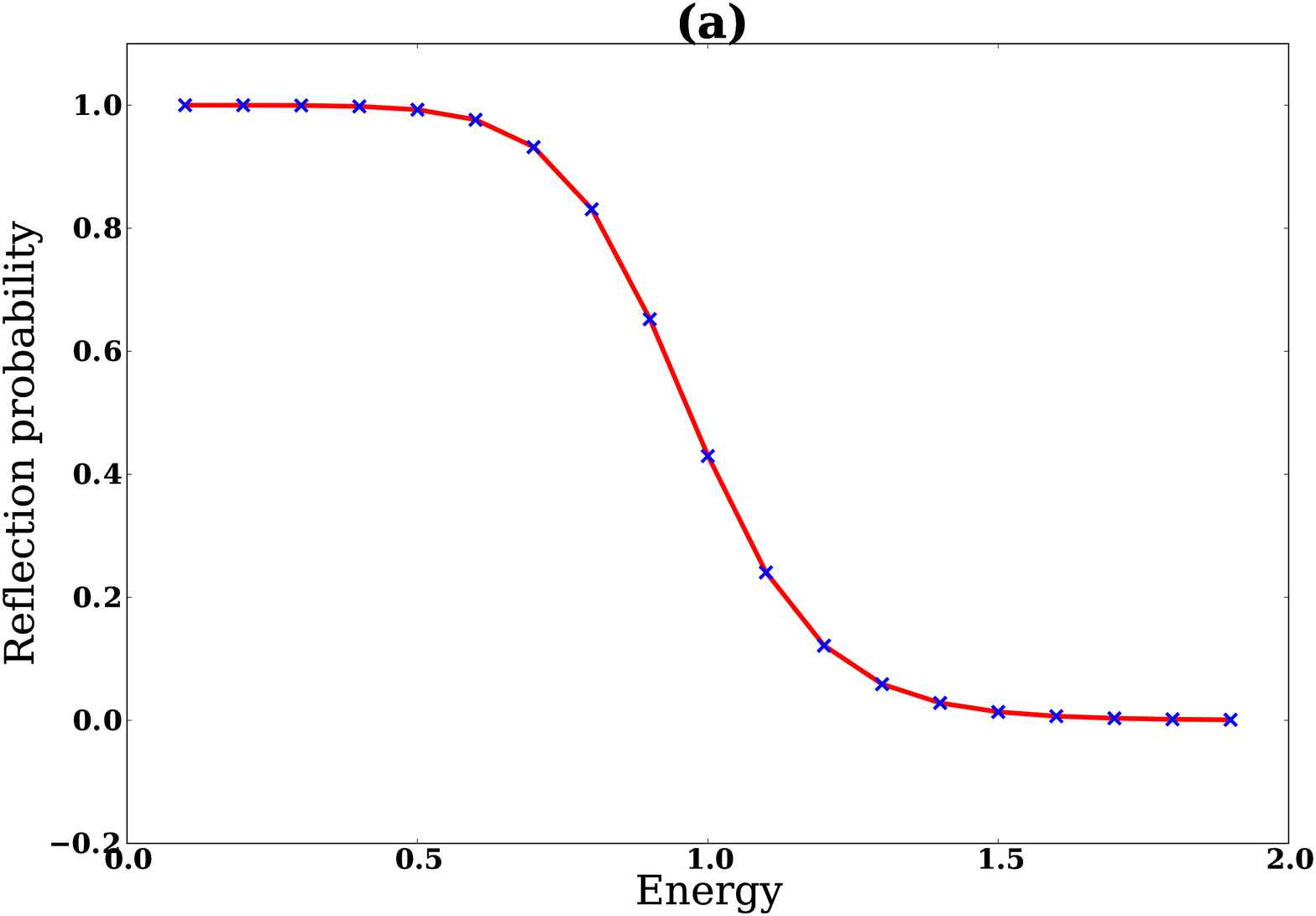}}, \resizebox {0.4 \textwidth} {0.3 \textheight }  
{\includegraphics {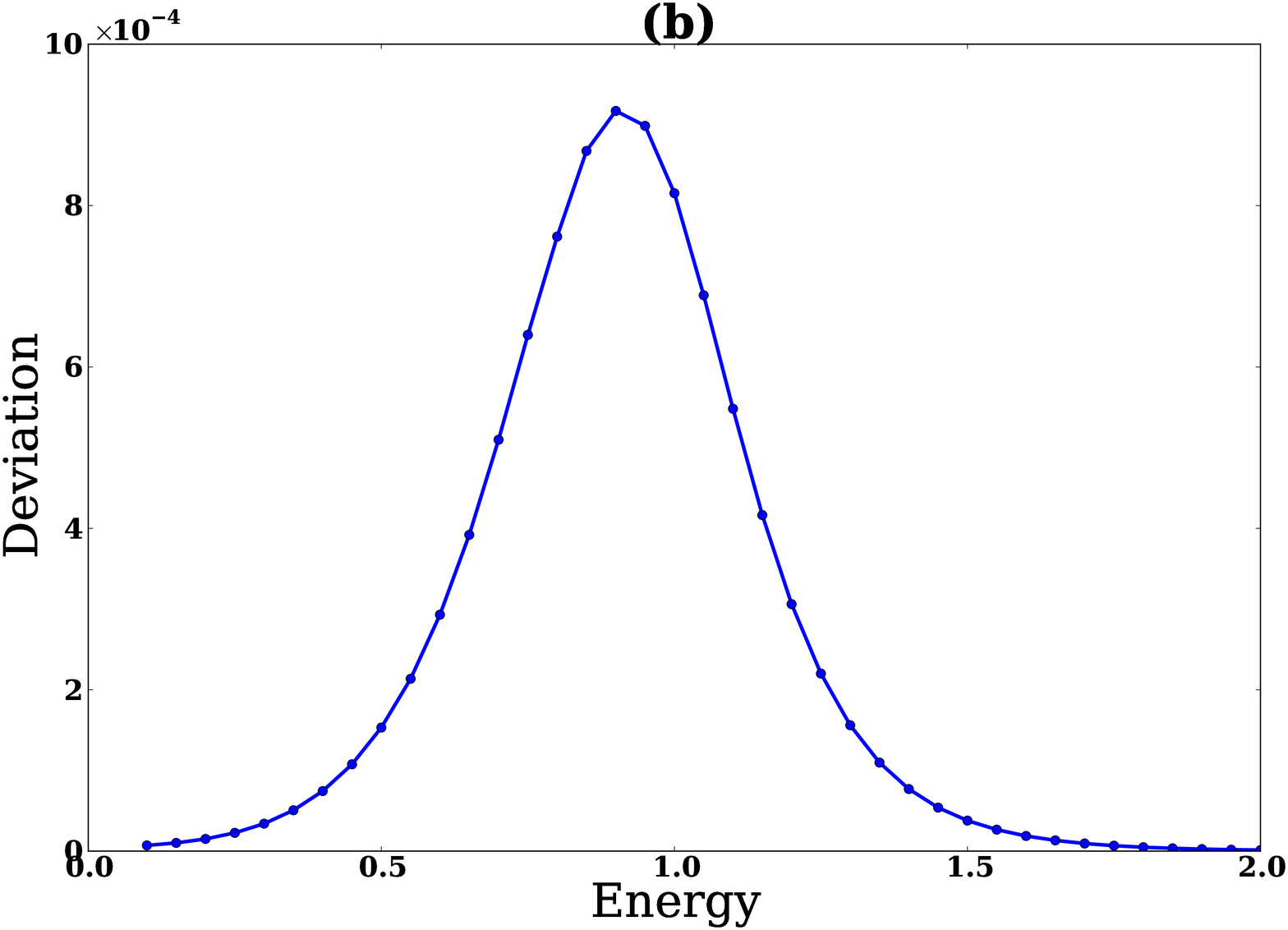}}
\caption{(a) Reflection probability  for symmetric Eckart potential with peak value $V_0=1$, for the energy values $0.1<E< 2$. Blue x marks indicates values obtained using ansatz (\ref{eq:ansatz}) and red continuous line indicate values obtained in standard quantum mechanics (b) Deviation of reflection probability with standard values,  for the energy  interval $0.1\leq E\leq 2$ of a symmetric Eckart potential (with peak value $V_0=1$).}  \label{fig:Eckart_prob_dev}
  \end{figure}

Using the ansatz (\ref{eq:ansatz}) to evaluate the reflection probability for the symmetric Eckart potential, with the period $2L=2\pi$, we observe that there is very good agreement with the standard result (\ref{eq:zafar_R}). The two graphs for the range $0.1<E<2$  are plotted together in Fig. \ref{fig:Eckart_prob_dev}(a). However,  slight deviations from standard result for moderate energies,  with the peak  of the  deviation reaching  $ 0.9\times 10^{-3}$  at $E=0.95$, is observed. [See Fig. \ref{fig:Eckart_prob_dev}(b).] This deviation is due to the periodicity of the pattern along the imaginary direction. The periodicity, with constant period,  is there for all $x_r$ and hence we cannot attribute the deviations to the choice of $\alpha$.  Since the ansatz (\ref{eq:ansatz})  has the special feature of being  independent of the assumption (\ref{eq:zafar_assmp}) and is based on the characteristic properties  of the trajectory pattern, we hold that the present results deserve further investigations. See that the maximum value of the deviation in this case amounts to only of the order of $0.1 \%$ of the actual  value. If the deviations are within the measurable range,   this prediction provides an opportunity to falsify the ansatz. 

\begin{figure} 
\resizebox {0.3 \textwidth} {0.3 \textheight }  
{\includegraphics {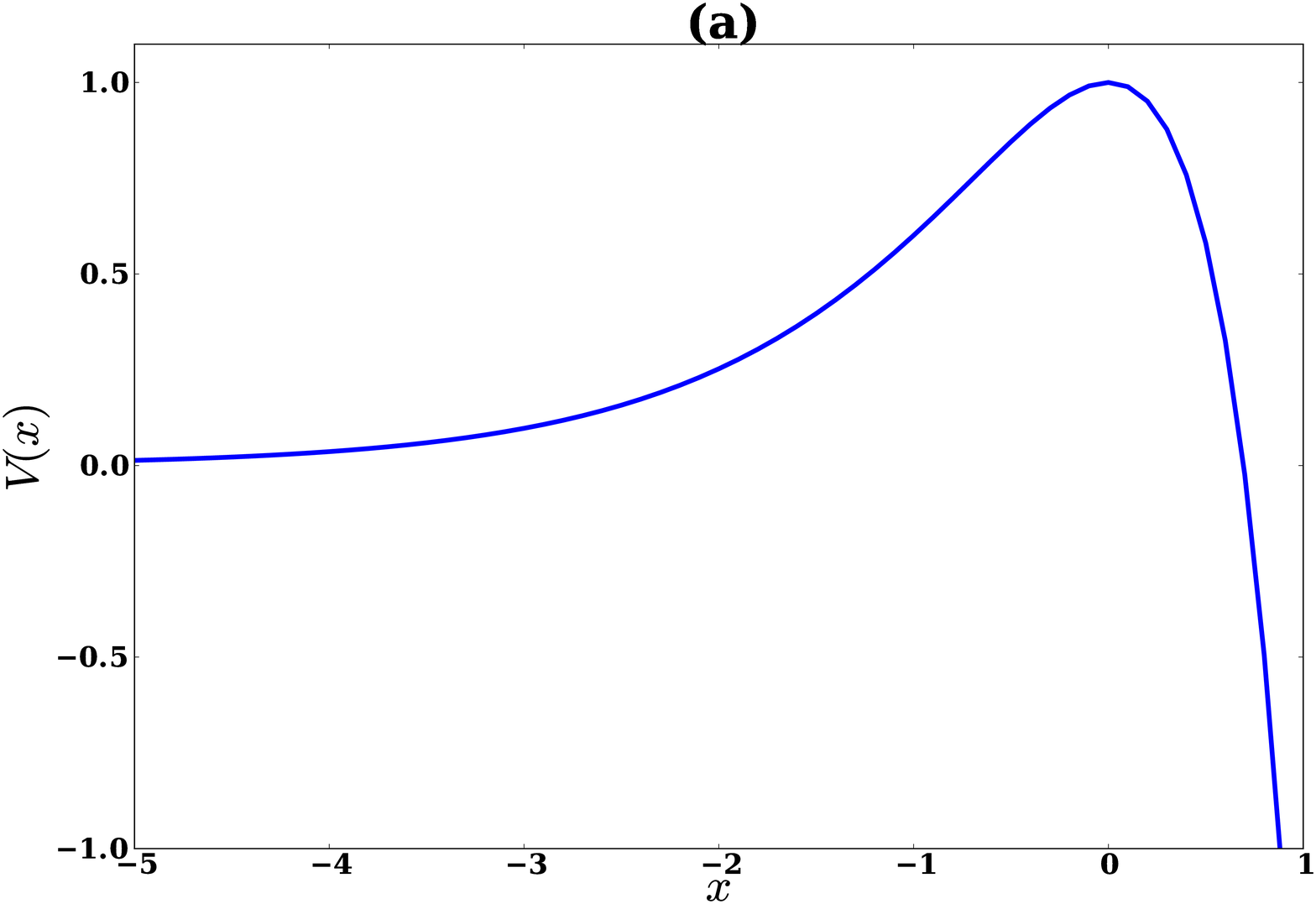}}, \resizebox {0.3 \textwidth} {0.3 \textheight }  
{\includegraphics {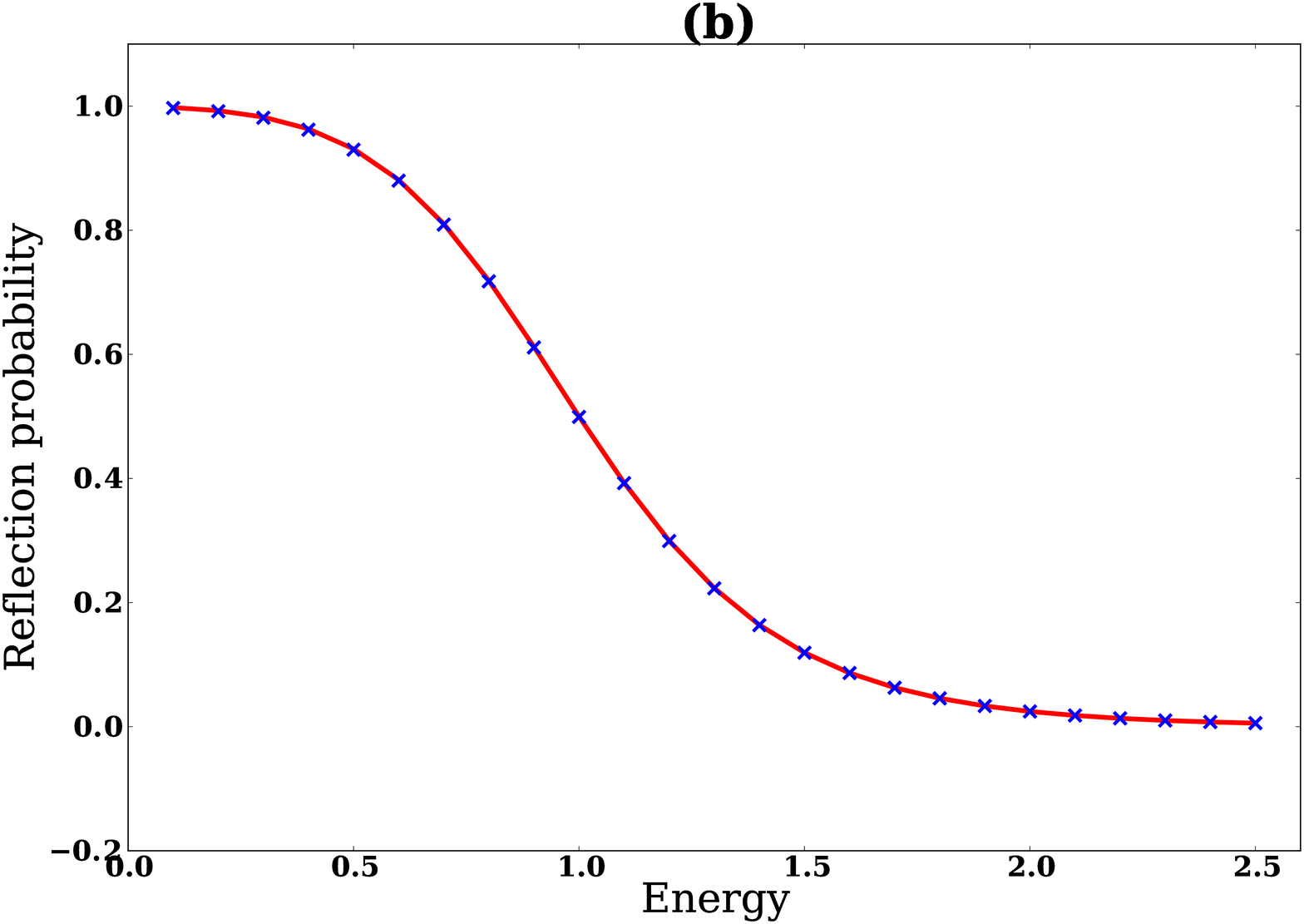}}, \resizebox {0.3 \textwidth} {0.3 \textheight }  
{\includegraphics {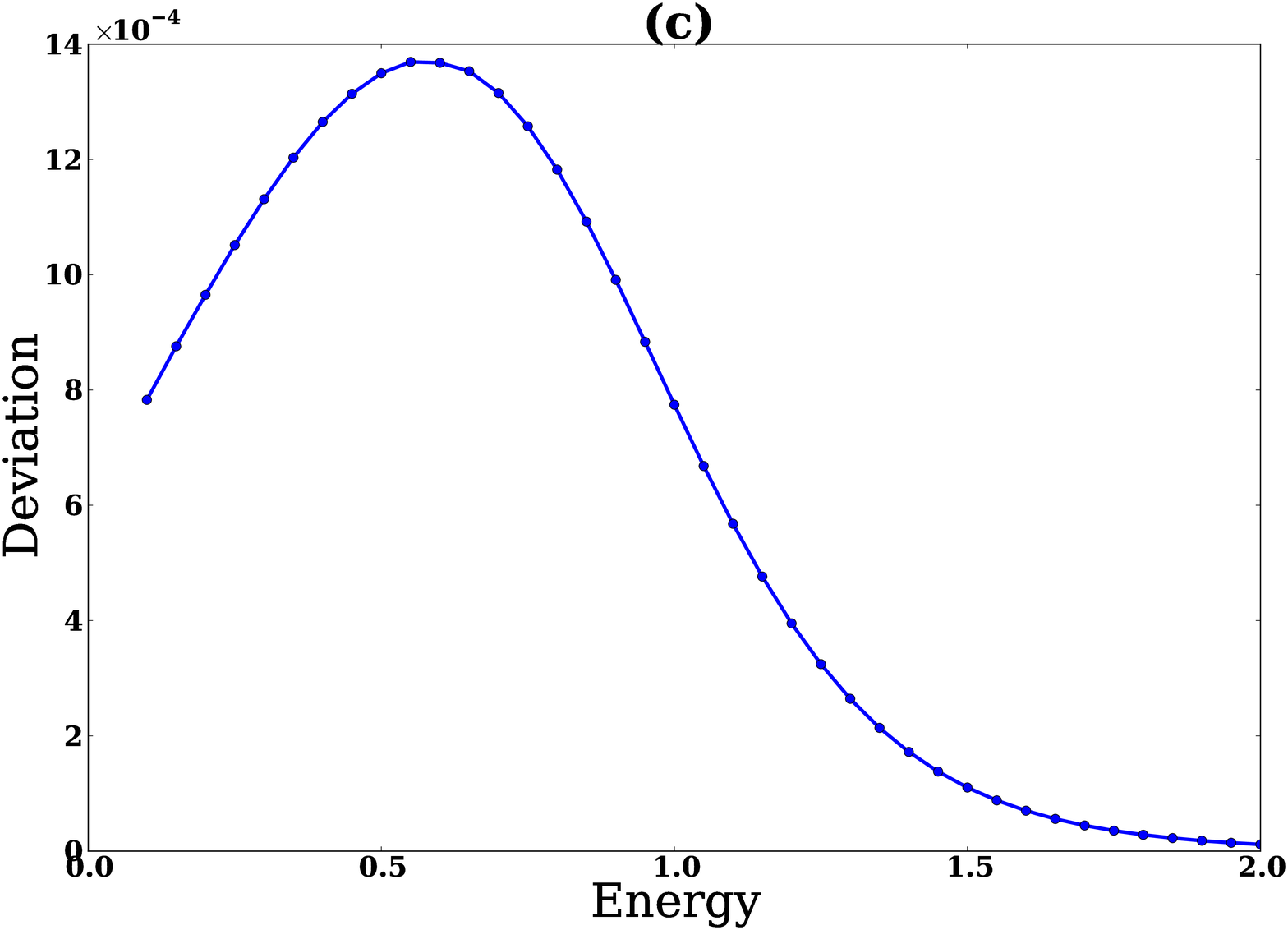}}  
\caption{(a) A potential very close to the Morse potential barrier (b) Reflection probability  for the above potential close to the Morse  barrier, with peak value $V_0=1$. Blue x marks indicate values obtained using ansatz (\ref{eq:ansatz}) and red continuous line indicates values obtained in standard quantum mechanics (c) Deviation of reflection probability  with standard values, for the above potential close to the Morse barrier.} \label{fig:Morse_potential_prob_dev}
  \end{figure}

Similar results for a potential close to the Morse potential barrier [with $c=0.001$ in  Ahmed's potential (\ref{eq:ahmeds_potential}) and depicted in Fig. \ref{fig:Morse_potential_prob_dev}(a)]  are obtained. The reflection probability versus energy, while using the MdBB approach, is plotted along with the standard results in Fig. \ref{fig:Morse_potential_prob_dev}(b). These are again found to be in very good agreement. As in the case of the Eckart potential, there are very small deviations for reflection probability in the case of moderate energies. Deviation of reflection probability  from standard values for this potential is given in Fig. \ref{fig:Morse_potential_prob_dev}(c).

\begin{figure} 
\resizebox {0.4 \textwidth} {0.3 \textheight } 
{\includegraphics {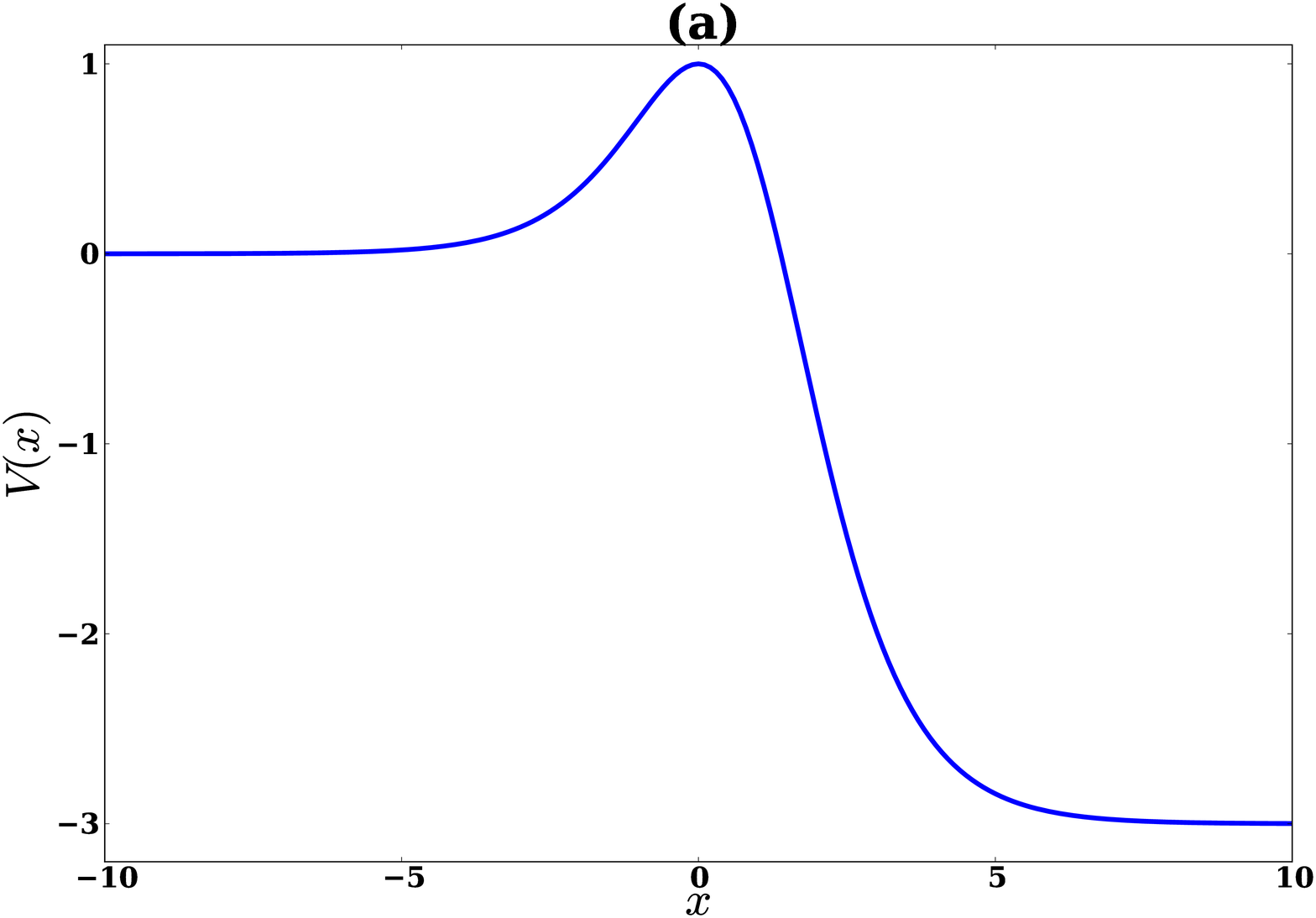}} ,\resizebox {0.4 \textwidth} {0.3 \textheight }  
{\includegraphics {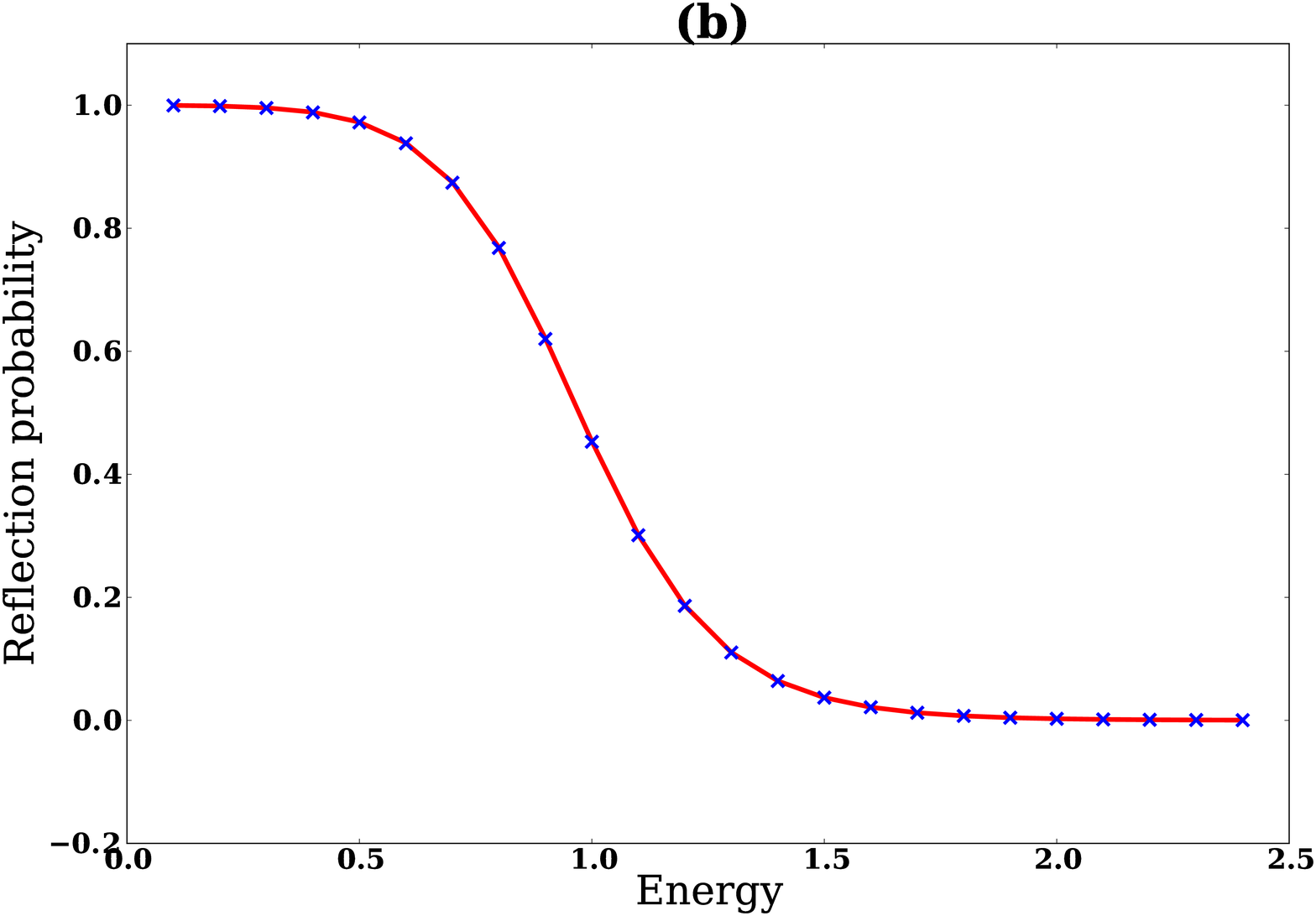}} 
\caption{(a) One-dimensional potential barrier (\ref{eq:ahmeds_potential}) with $c=0.5$ (b) Reflection probability  for the potential barrier (\ref{eq:ahmeds_potential}), with $c=0.5$ and peak value $V_0=1$. Blue x marks indicate values obtained using ansatz (\ref{eq:ansatz}) and red continuous line indicates values obtained in standard quantum mechanics.}  \label{fig:mid_potential_prob}
  \end{figure}

  Another potential midway between the Eckart and the Morse barriers [with $c=0.5$ in  Ahmed's potential (\ref{eq:ahmeds_potential}) and depicted in Fig. \ref{fig:mid_potential_prob}(a)] also shows excellent agreement with standard results, for low and high energies, as can be seen from Fig. \ref{fig:mid_potential_prob}(b).

\section{Transmission Over the Soft Potential Step}

\begin{figure} 
\resizebox {0.4 \textwidth} {0.3 \textheight }  
{\includegraphics {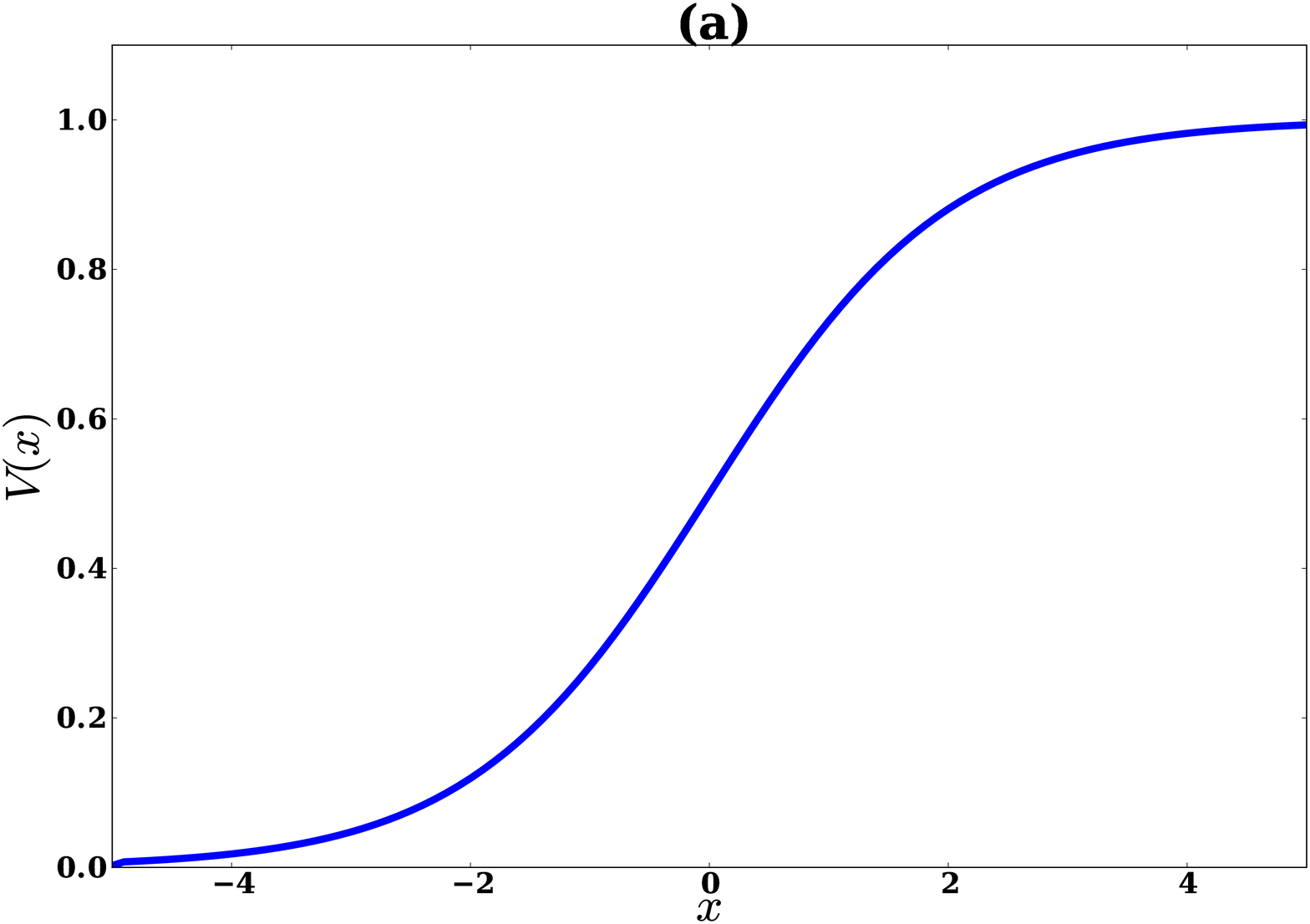}},\resizebox {0.4 \textwidth} {0.3 \textheight }  
{\includegraphics {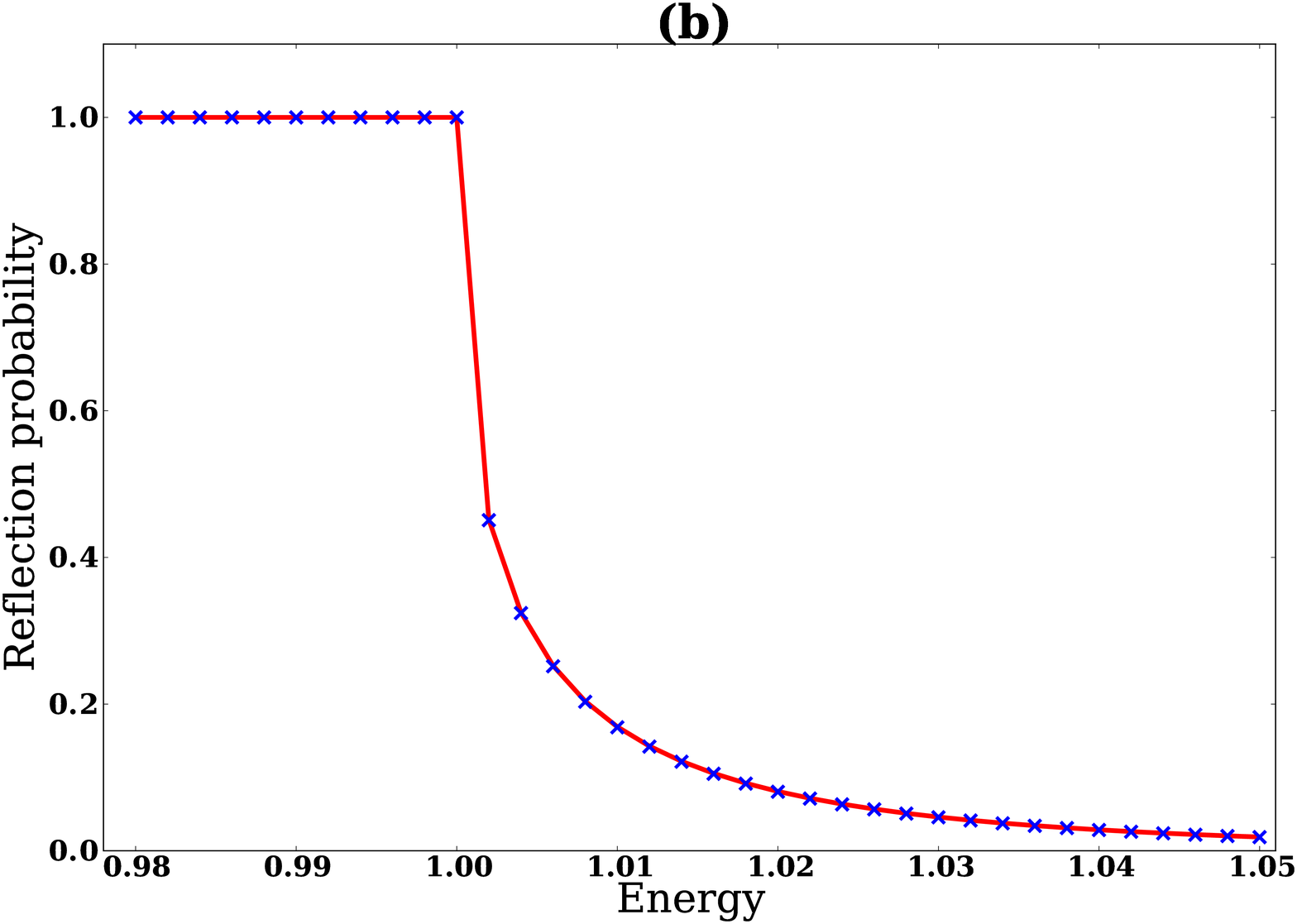}} 
\caption{(a) Soft  potential step (b) Reflection probability  for soft potential step with step height $V_0=1$. Blue x marks indicate values obtained using ansatz (\ref{eq:ansatz}) and red continuous line indicates values obtained in standard quantum mechanics.}  \label{fig:Soft_potential_prob}
  \end{figure}

\begin{figure*}  
\resizebox {0.33 \textwidth} {0.3 \textheight }{\includegraphics {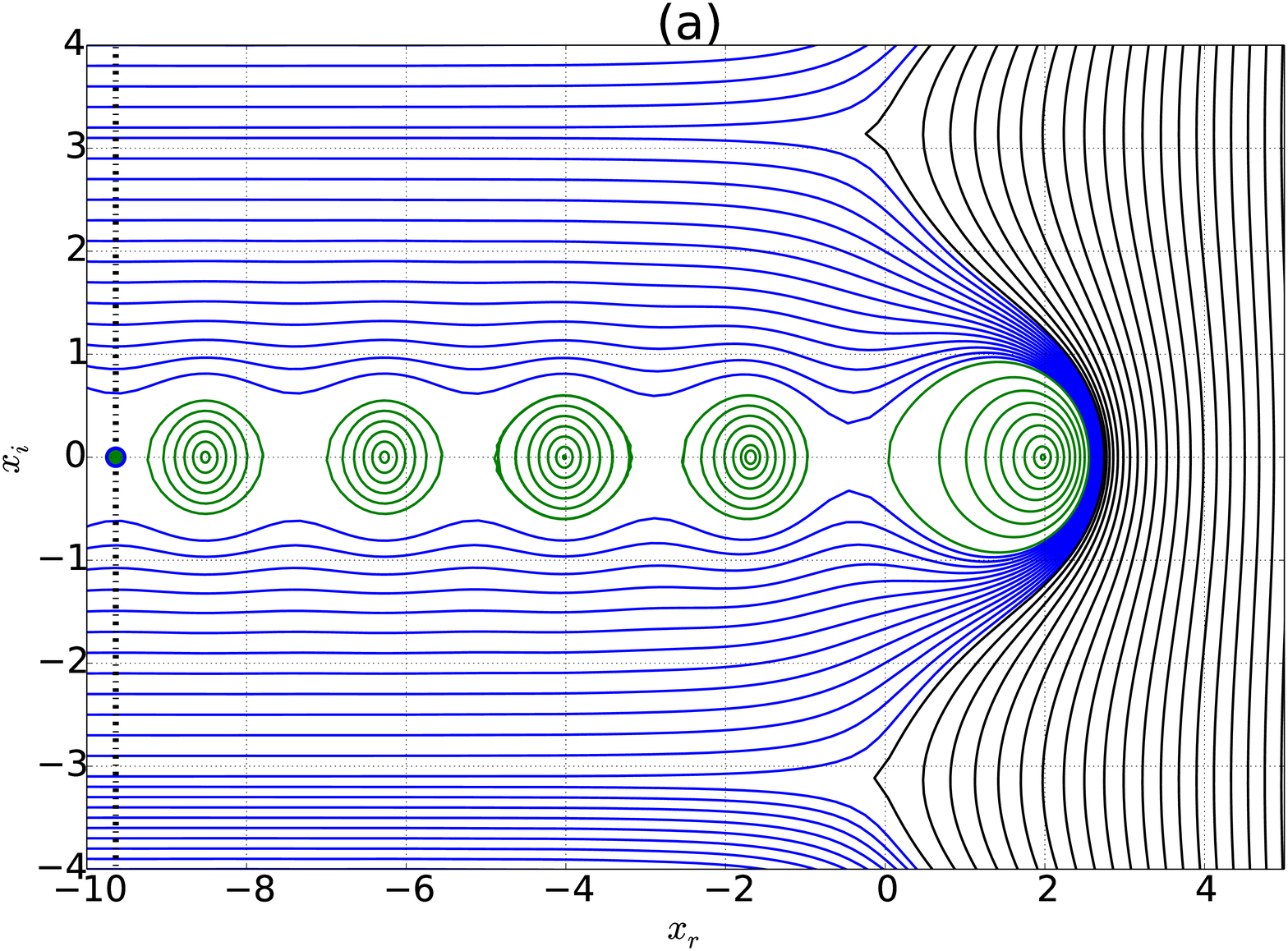}},\resizebox {0.33 \textwidth} {0.3 \textheight }{\includegraphics {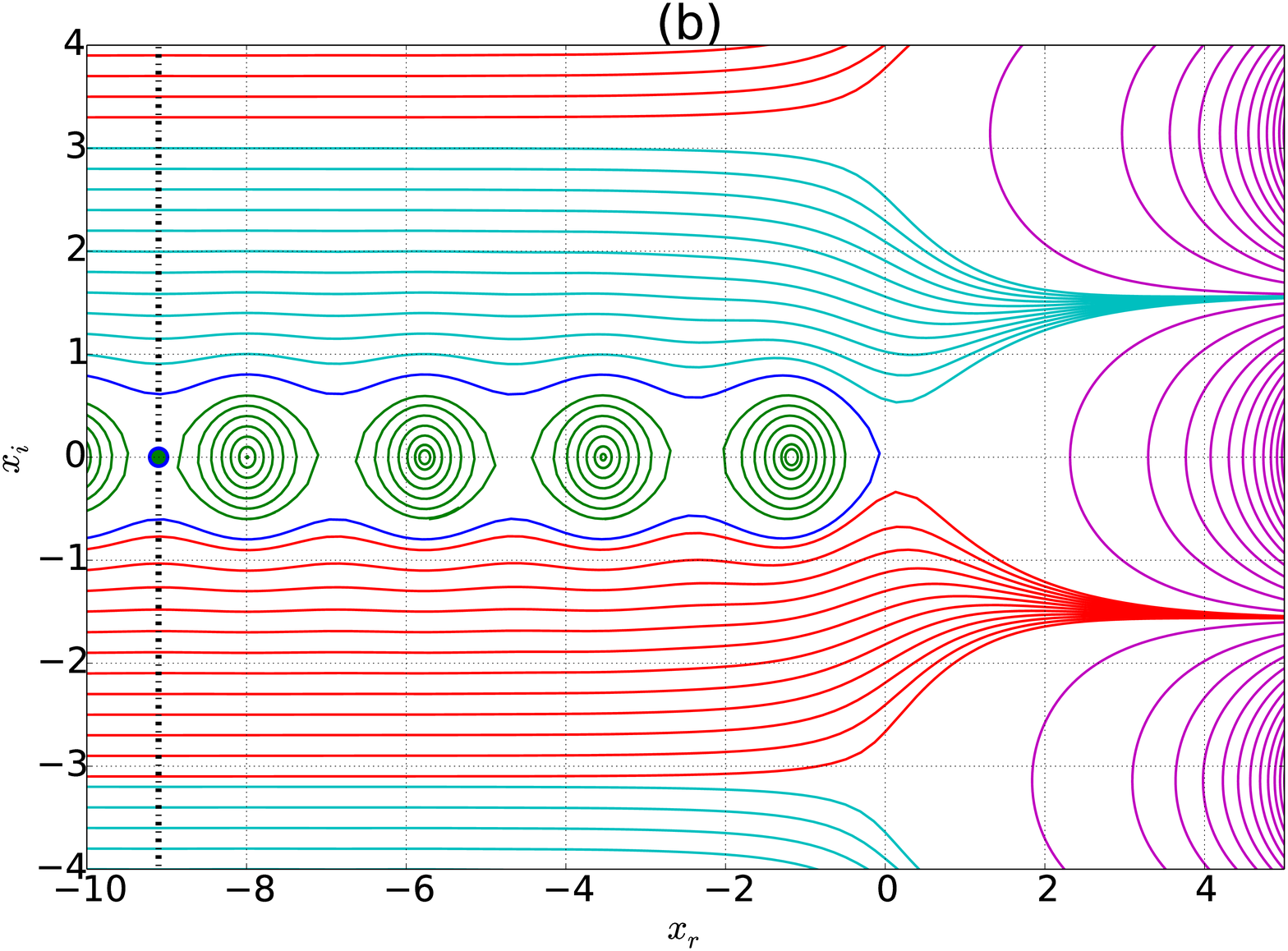}},\resizebox {0.33 \textwidth} {0.3 \textheight }{\includegraphics {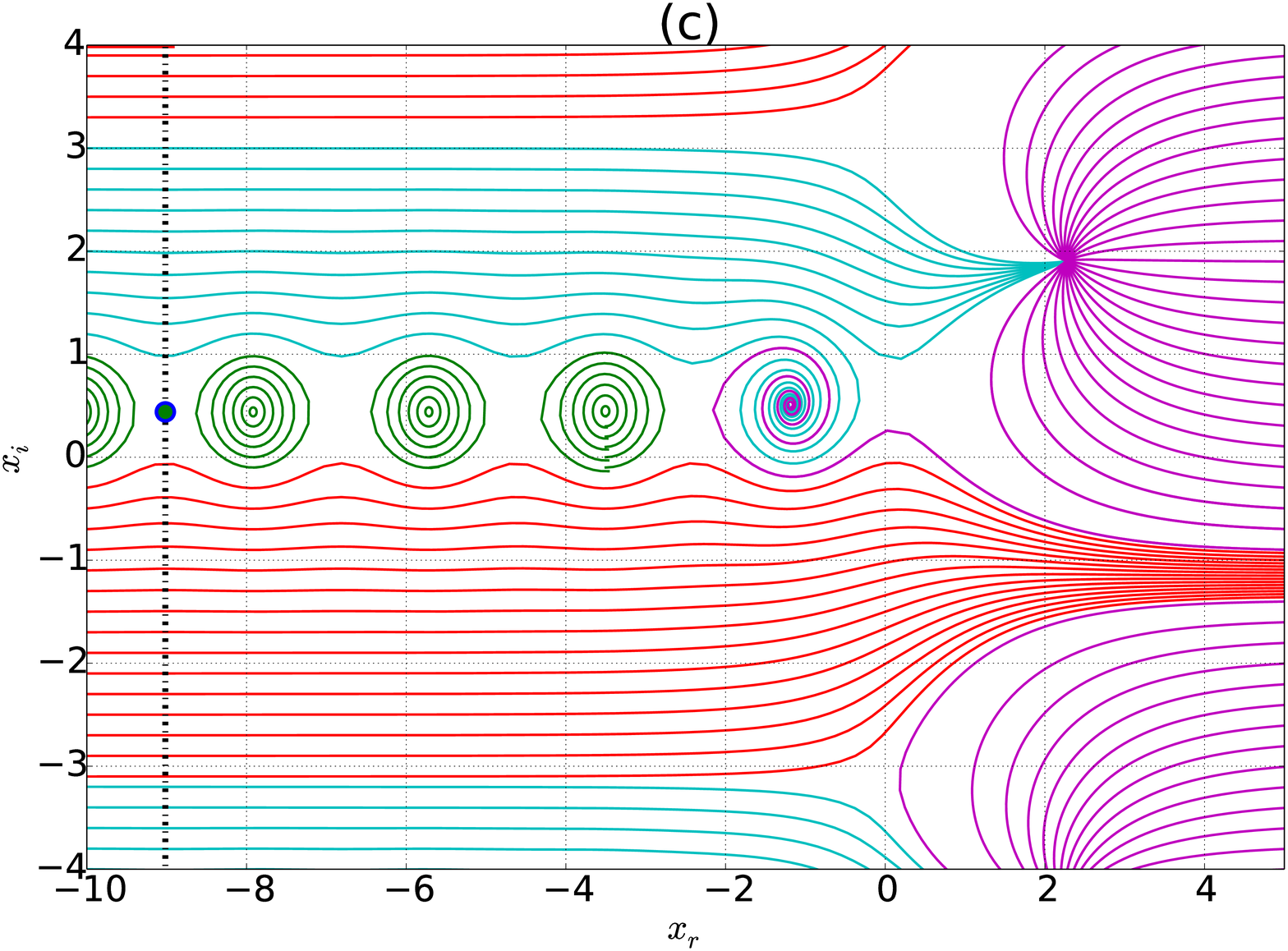}} 
\caption{Complex quantum trajectories for soft potential step, for the energy values (a) $E=0.98$, (b) 1.0 and (c) 1.02. Color codes are as in Fig. 1. In each panel, the points $\alpha +i\beta$, whose numerical values  are $-9.63772+	0 \; i$, $-9.1014+ 0\; i$, $ -9.01653+	0.440077\; i$, respectively,  are marked with $\bullet$ symbols. }  \label{fig:Soft_trajs}
  \end{figure*}

The complex quantum trajectories for a rectangular potential step  \cite{mvj1} show no periodicity. Hence for evaluating the reflection probability in this case, one needs to use the ansatz (\ref{eq:ansatz_2Lambda}). We have seen that when this is worked out, there is perfect agreement with standard result, as $\Lambda \rightarrow \infty$.  In this section, we turn to the evaluation of the probability for a soft potential step, which shows  periodicity along the imaginary axis.  
Thus we consider a soft potential step of the form \cite{flugge},

\begin{equation}
V(x) = \frac{1}{2} V_0\left( 1+ \tanh \frac{x}{2a} \right). \label{eq:soft_potential}
\end{equation}
The potential $V\rightarrow 0$ as $x \rightarrow -\infty$ and $V\rightarrow V_0$ as $x \rightarrow \infty$, as shown in Fig. \ref{fig:Soft_potential_prob}(a). Since they contain a factor $\exp(x/a)$, also these potentials have periodicity along the imaginary direction, as in the case of Ahmed's potential discussed above. Using  $y=(1+e^{x/a})^{-1}$ instead of  the variable $x$, the solution to the time-independent Schrodinger equation can be written in the form \cite{flugge}

\begin{equation}
u(y)=y^{\nu}(1-y)^{\mu} f(y).
\end{equation}
Here 

$$
\nu^2=\frac{2ma^2}{\hbar ^2} (V_0-E), \qquad  \mu^2=-\frac{2ma^2}{\hbar ^2}E,
$$
and $f(y)$ is given by

\begin{equation}
f(y)=C\; {}_2F_1(\mu + \nu,\mu +\nu +1, 2\nu +1;y),
\end{equation}
where the constant $C$ is to be fixed to satisfy the boundary conditions. However, the present MdBB approach does not need the value of $C$ to evaluate the reflection probability. Instead, with the value of $x_i=\beta$ corresponding to a pole for $\dot{x}$  as described above, we can  evaluate $R(k)$ by using the ansatz (\ref{eq:ansatz}). Complex trajectories for the soft potential step, for values of energies $E=0.98 V_0$, $E=1.0 V_0$, and $E=1.02 V_0$, respectively, are plotted in Fig. \ref{fig:Soft_trajs}. Here also, the trajectory pattern shows periodicity along the imaginary axis, with period $2L=2\pi$.

The reflection probability $R(k)$ obtained using the present approach and that using the standard approach \cite{flugge} are plotted together in Fig. \ref{fig:Soft_potential_prob}(b), for the values of $V_0=1$ and $a=1$. The agreement is very good in this case too, but  when $V_0$ is very small, the situation can change and one may find deviations from standard results, as in the case of Eckart and Morse potentials. Another  peculiarity in this case is also worth noting. The potential (\ref{eq:soft_potential}) shows  periodicity, with period $2\pi a$. As the step becomes sharp with $a\rightarrow 0$, the period will be very small and consequently deviations from standard result may increase. But this  type of a sharp potential step, with very small period,  is different from the rectangular potential step, which has no periodicity at all \cite{mvj1}. For the soft potential step,  when either $a$ or  $V_0 $ is very small, there can be  deviations from standard results.

\section{Conclusion}

The present work is in continuation of a previous study of quantum trajectories in complex space  \cite{wyatt_eckart,wyatt_eckart1}, where among other results,  the exact complex quantum trajectories for the Eckart  potential barrier and the soft potential step were obtained. Plotting the complex trajectories by integrating  the  equation of motion $\dot{x}\equiv ({1}/{m}) {\partial S}/{\partial x}$, it was  shown in their work that more trajectories link the left and right regions of the barrier, when the energy is increased. In the present paper, we attempt to  evaluate the reflection and tunneling probabilities on the basis of these observations. It is true that these probabilities are obtained if one goes back to the standard procedure in  quantum mechanics, and evaluate them using the $\psi^{\star}\psi$ distribution along the real line. But we argue that  one can take the complex trajectory representation and the complex-extended probability density seriously only when it is possible to evaluate the reflection and tunneling probabilities  with the help of these complex entities. Another clear motivation for the work is that  the MdBB complex trajectories themselves suggest  this possibility. 

We have also noted that  issues such as tunneling time remain as open problems in    standard quantum mechanics. On the other side, in the  tunneling of particles in energy eigenstates, the dBB trajectories give an unphysical  prediction that all incident particles tunnel through the barrier.  But as in all other predictions of observable physical quantities, the  dBB approach adopts the same probability evaluation scheme  used  in standard quantum mechanics and claim equivalence with its results.    In this paper, we investigate this issue and show that the new MdBB scheme can correctly describe the tunneling of particles in energy eigenstates and that the evaluation of reflection/tunneling probabilities can be done  with the help of the complex-extended probability density and the trajectory pattern.  The reflection probability is found as the ratio between the total probabilities of the reflected  and the  incident trajectories, the latter quantities being obtained by integrating the  complex-extended  probability density along the imaginary direction.  

         The calculations are performed  for a rectangular potential barrier,  symmetric Eckart and Morse barriers, and a soft potential step.  It is easy to see that for the rectangular potential barrier, the agreement of  $R(k)$ in Eq. (\ref{eq:ansatz_2Lambda})  with the standard result $  |B|^2/|A|^2$ as $\Lambda \rightarrow \infty$ shall be exact.  In the case of other smooth, realistic potentials,  there are slight deviations from standard results 
 for moderate energies ($E\approx V_0$). These deviations arise from the periodicity of the trajectory pattern along the imaginary direction. (Since the periodicity, with constant period, is there for all  $x_r$, the deviations cannot be attributed to the choice of   $x_r=\alpha$, the line along which the integration is performed.)    The results obtained are in very good agreement with   the standard results and the  predicted deviations are very small,  with  a maximum value  only as much as $0.1 \%$ of the original  value. This validates the proposed complex-extended quantum probability density proposed in \cite{wyatt_prob1,wyatt_prob2,mvj_prob2}. Measurement of these small deviations shall provide an opportunity  to falsify  the ansatz.      The new ansatz is preferred, for it is based on the   characteristic feature   that  incident   trajectories  either get reflected or  transmitted and that more trajectories link the left and right regions of the barrier when the energy increases.    Hopefully, the new approach  will  lead to wide-ranging applications of the complex trajectory representation of  quantum mechanics. 
 
 \medskip
 
\noindent {\bf Acknowledgements} 
 
\medskip 
 
\noindent The authors wish to thank Prof. K. Babu Joseph and Dr. Zafar Ahmed for valuable discussions and KM wishes to thank the Santhom Computing Facility, Kozhencherry,  for hospitality.

\end{document}